# ANALYSIS OF USING BROWSER-NATIVE TECHNOLOGY TO BUILD RICH INTERNET APPLICATIONS FOR IMAGE MANIPULATION

by

Thomas Steenbergen

A Thesis Submitted to the
Department of Mathematics and Natural Sciences
in Partial Fulfillment of the Requirements for the Degree of

MASTER OF COMPUTER SCIENCE

© Thomas Steenbergen, 2009
University of Leiden

# Acknowledgements

First and foremost, I want to thank all those who assisted me in the creation of this thesis.
I would especially like to thank all my former and current colleagues at Joost Technologies who inspired and helped me in various ways into becoming a proper software engineer. Of all my colleagues I want to express my special thanks to Rick van der Zwet for his guidance during the project and Henk Uittenbogaard for providing an excellent and inspiring working atmosphere.

I would also like to thank Dr. Michael Lew, my examiner at LIACS, Leiden University, who has provided me with constructive feedback during the realisation of the web application and the throughout the creation of this thesis.

Finally, I owe special gratitude to my parents and family for their continuous and unconditional support enabling me to accomplish this milestone in my academic career.

# Table of Contents

## Table of Contents (cont.)

**LIST OF FIGURES**

**LIST OF TABLES**

# Abstract

The majority of the well-known web applications are using are using browser-native web technology like HTML and Ajax in their rich internet applications to provide email, spreadsheet and word processing to their customers. The situation however changes when it comes to online image manipulation since most of those websites are using browser plug-in based solutions like Adobe Flash. This raises the question of why these sites switch to a proprietary format whilst there are multiple browser-native technologies like Canvas, SVG and VML that on paper can be used to display and manipulate images.

In this thesis we will investigate whether browser-native technologies can be used to perform photo manipulation tasks e.g cropping, resizing or rotating an image within the current mainstream browser. By the use of a case study we will analyze problems that have occurred during the implementation of a prototype web application that utilizes browser-native web technology in order to create an online version of a real world photo scrapbook. Implementation of a prototype will allows us to analyze the strengths and weaknesses of current web technology when it comes to browser-based image manipulation.

Furthermore we explore the possibilities of the Ajax in combination Canvas, SVG and VML to provide a more interactive graphical user interface to perform image manipulation tasks on the web.

Keywords: Ajax, browser-native, Canvas, Framework,  JavaScript, JSON, Rich Internet Application, SVG, VML, XML, web application, web development, web standards

# 1. Introduction

## 1.1 Background

The idea that forms the basis of writing this thesis was a frustration which occurred when we wanted to crop and rotate an image but we didn't have sufficient administrator rights on the computer where we were working on to install new software.

As being part of the internet generation we immediately thought of using one of the many online photo editing applications to perform the task, however we were unable to proceed since almost all of them required a browser plug-in technology called Adobe Flash [1]. It was not the case that we did not have the software installed but most websites simply denied access since the computer we used did not have the right version of the software installed.

This highlights one the many disadvantages of using browser plug-in technology and after thinking and researching the subject we found out that there are actually quite some valid reasons why one should opt not to use browser plug-in technology like Adobe's Flash to offer web services:

**Accessibility** - Flash animations in most websites almost never come with keyboard navigation whilst some users are just not able to use a mouse. Forced by Americans with Disabilities Act (ADA) which passed the United States Congress in 2001 Adobe has been working hard to get their technology more accessible [2]. However screen readers only exist for the Windows operation system and their performance is reported to be mediocre in combination with Flash.

**File size** - DHTML and JavaScript files are simple small text files in contrast to Flash files which are always bigger since they are actually pre-compiled executables. Furthermore one does not have the option when browsing the web on a mobile device to leave out the images on a Flash powered website. Looking from the perspective of a large web application provider like Google or Yahoo with millions of users every day then size is everything: the smaller the application, the more money is saved on bandwidth costs.

**Poor degradation** - Application degradation means that even if the user does not have the right version of piece of software still he should be a view part of the application or the data within it. Flash websites usually have a simple policy on degradation either you have a particular minimum version of the software installed or nothing will be displayed.

Degradation has been a big issue especially for users of the Linux operating system. It took Adobe systems Inc., the maker of Flash until December 2008 to released version 10 which was the first universal version across all mainstream operating systems. Before that release if you were using the Linux or Apple operating system then sometimes your browser would not be able to display Flash content due to fact that the latest version of Flash was not available for your operating system.

**Poor performance** – A well-known complaint from users is that the Flash version for Linux and Mac is slower and uses more computer resources when playing the same content than the Windows version. Overall the displaying of Flash websites and movies is very CPU intensive and many users complain that Flash is the main reason for their web browser too crash when browsing the Internet.

**Proprietary standard** - Although the FLEX SDK [3] has been open-sourced, the browser plug-in a.k.a. the Flash player and its IDE are not, meaning the source code is not available for all to read, change and redistribute.
This goes against the open architecture of the Internet in which so called Requests For Changes (RFCs) [4] are used to determine protocols and standards. These RFCs are available for everybody to read and you are invited to join discussions on new standards. It is believed by many that this open character of the Internet will stimulate its growth because it allows the actual specifications to be used in college classrooms or by entrepreneurs developing new systems.

**Search Engine Friendliness** - Basically every flash site on the World Wide Web is currently not indexable by every major online search engine. Search engine spiders, which are little programs that index the worldwide web, face the same issue as screen readers for people with disabilities; they are unable to access and read to information within the webpage since it is stored in an executable. Most of the major search engine service like Google therefore advice flash websites to include a secondary HTML webpage with a description of what's on the website so it can be found by their customers.

In the above list we focused on Flash since it is the most widespread browser plug-in technology but most of the reasons can also be applied to other browser plug-in technology like Sun Microsystems's Java Webstart and Microsoft's SilverLight.

The long list of disadvantages and the rise in popularity of online application websites got us thinking: Google Mail [5] and Maps [6], Yahoo! Mail [7] and Zoho [8] are using browser-native technology like Ajax [9] to give consumers the ability to work on their mail, maps, documents, presentations and spreadsheets. Why are then almost all of the online photo editing websites using Adobe Flash? Is it not possible to use browser-native technology to design a simple web application that can perform simple task as image cropping, resizing and rotating? What are the difficulties involved in building such an application?

All these questions inspired us to research web-native technologies and write this thesis.

## 1.2 Problem

In recent years many of the classical desktop applications like email clients and the standard office suite (word-processing, spreadsheets etc) have been turned into web applications [10]. This move toward online applications has been made possible with the increase of affordable high-speed Internet connections which enabled people to have an 'always on' relationship with the World Wide Web.

Switching applications to the web has several advantages for one it's a low-cost and effective way to distribute software to millions of users. New functionality can be added and deployed within minutes making it easier for organizations to update their services. On the other end there is no world as complex as developing for the Internet were developers have to mix multiple programming languages, protocols and applications to get a working web application [11].

Providers of web applications who have millions of users prefer using browser-native text-based mark-up languages to build rich user interfaces since it generates far less web traffic than plug-in based applications and thus saving them a lot of money in broadband costs. Plug-in based technologies like Flash or Silverlight do have their benefits when working with multimedia content. It is relatively easy to incorporate browser-based image manipulation with numerous features into a flash-based web application. Most of the browser-native web applications with image editing capabilities usually only offer a very limited set of functionality e.g. resize, rotate 90 or 180 degrees since for every manipulation a roundtrip to the image processing backend servers is needed. Large photo community websites like Flickr [12] want to give their customers a good set of image editing tools but also save money on bandwidth costs. That's why they require you to install a browser plug-in like Adobe Flash in order to use their photo editing web application.

What most people do not know is that the average browser supports a handful of languages that on paper can also perform image manipulation tasks. The question that now arises is why has there not been any commercial party that uses these browser-native languages to build a photo editing web application. What are the real-life disadvantages of current browser-native technology that is stopping them in releasing such an application?

## 1.3 Purpose

The purpose of this thesis was to research current browser-native technology and how they can be applied to online photo manipulation. We are not going to perform a comparative study of browser-native versus browser plug-in technology e.g. Ajax versus Adobe's Flash.

We are more interested in trying to answer the following questions:

- Which browser-native standards that are capable of performing image manipulation tasks are widely supported by the mainstream web browsers?

- What are the challenges or roadblocks facing web developers when using browser-native technology to build a web application?

- What are the rewards of using browser-native technology in a photo web application for instance how can they be used to enhance the user experience?

## 1.4 Methodology

The basis of thesis will be case study of current browser-native web technology and how it can be applied to perform client-side photo manipulation.
A case study has been defined by Robert K. Yin [13] as:

*A case study is an empirical inquiry that*
*- investigates a contemporary phenomenon within its real-life context especially when*
*- the boundaries between phenomenon and context are not clearly evident*

This means that the subject of study can only investigated in its natural context since the context is pertinent to the subject. Technologies that make up the World Wide Web are rapidly changing or evolving and they are also all interconnecting or supporting each other. Therefore we believe browser-native technology can only be research by performing a case study.

We will document any problem that arises whilst implementing a prototype web application that incorporates several of the available technologies that according to their specifications or according to available literature can perform client-side image manipulation.
There is a significant challenge since in order to properly compare technologies we have to implement the same web application functionality in each of them. On completion of the programming of the web application the used technologies will analyzed on several aspects which will include performance, stability and usability.

As we have already mentioned, technologies used to create web applications are rapidly changing or evolving which has as a consequence that the traditional academic cycle of publications are unable to keep up. In order to keep up with the pace of development the original academic developers switched to more informal requests for comments. This has as result that there are not a lot academic publications in our field of research especially papers on web standards and browser technologies are hard to come by. Nevertheless we have strived to only reference to relevant literature that was made available to us through the libraries of the University of Technology Delft and the University of Leiden.

If no academic verified literature was at hand on a particular topic then we have resorted to books about web technology in general with chapters or paragraphs describing a particular language or standard. In this thesis we explicitly did not include any references to the Internet due to the volatile nature of this medium. We believe our reader is quite capable to use an online search engine to find the official web pages of some of the technologies and products we discuss within this thesis.

## 1.5 Limitations

There are multiple web browsers available to consumers however all of them support a different range of web technologies therefore for this thesis we have decided that we focus on the big four mainstream browsers, namely Microsoft Internet Explorer, Mozilla Firefox, Opera and Apple's Safari. We will only cover technologies in this thesis that are out of the box supported by the above named browsers.

Although recently Google released its own web browser Chrome which has according to some market analysis taken over Opera in the number of users, we found out that it is still in heavy development and hence it will be ignored during the implementation of our prototype web application and our experiments.

We also have to account for that fact that there are multiple versions of the same browser since software is in general periodically updated with a new release and version number every couple of months e.g. Internet Explorer 6, 7 and 8. This would extend the scope of this thesis too much in our opinion hence we restrict the set of browsers to the following versions: Internet Explorer 8, Firefox 3.5, Safari 4 and Opera 9.6. (for exact version numbers see paragraph 4.1.2).

There is also a vast array of programming languages and technologies that can be used to develop a framework that is needed to power the server backend of any web application. In order to make it easier for the community to extend and use the web application that we have developed we thought it was best to use the well-known AMP stack [14]. This combination of Apache, MySQL and PHP (AMP) is due to its open-source nature widespread, well-documented and is relatively cheap to set up and operate.

## 1.6 Structure

The first chapter of this thesis is an introduction chapter that provides the reader an insight in the reasons why this thesis was written. It discusses the problem that we are trying to solve and how we will approach it. Also includes the restrictions that have been set to limit the scope of this thesis. In the second chapter we hope to familiarize the reader with the basics of several web technologies relevant to this thesis and in the third chapter we will discuss the inner-workings of the prototype that was developed to answer the main research question. Chapter 3 will be entirely devoted to the implementation of the prototype and the problems we encountered during development. Chapter four is where we have written a description of all the experiments that have been performed to see how well the prototype works in real-life. In the chapter five we analyze the experiments and discuss the results after which the final chapter will answer our main research question and look at future works.

# 2. Preliminaries

Throughout this thesis many different web technologies are mentioned and we understand that it might be confusing to the average reader on how they all work together to create a single web application. Therefore we have written a chapter that we hope will give the reader a better understanding of the basics of how an average web page is built up. We only discuss every technology briefly and we would like to encourage you to have a look at the included references if you desire a more complete and in-depth insight into a particular technology.

## 2.1 The building block of a webpage

### 2.1.1 HTML.W3C, CSS and DOM

In the beginning of the internet, there was nothing else than a simple collection of HTML documents which were linking to each other using hyperlinks. HyperText Mark-up Language (HTML) [15] was originally developed by Tim Berners-Lee and based on SGML (Standard Generalized Mark-up Language) [16] which is an international standard for marking up text into structural units such as paragraphs, headings and list items. Its creation came from his frustration that data was locked into different incompatible formats whilst at CERN [17] and therefore a new language was needed that was independent from the formatter e.g. the browser and was small enough so it could easily be exchanged.

It did not take long before others mainly academics to see the potential of Tim's invention and started contributing to expand and improve the language. The World Wide Web Consortium (W3C) [18] was formed to fulfil the potential of the Web by attempting to develop and keep web standards open so everyone can freely use and contribute back, aiding to its overall development. If a technology becomes a W3C standard its future development will be managed by the W3C consortium. The original creator of the technology has to license it royalty-free available to any person or organization that wants to implement and develop applications using the technology.

Along the way Cascading Style Sheets [19] came about as a W3C standard to set the style and look of a document written in a mark-up language. CSS enables you to separate presentation of a document e.g. colours and fronts from its contents. This separation makes it easier to re-use the contents of a document by for instance screen readers which make it accessible by voice or Braille to people with disabilities. It also facilities a rapid change of a website's look and feel without having to touch the contents.

The Document Object Model [11, 20] is language-independent convention for representing and interacting with HTML objects within a webpage. All the data a web browser gets from a server when retrieving a web address is stored internally within the browser into the DOM tree and is then displayed to the user. We explicitly mention DOM here since all web standards that an average browser supports are all attaching onto or manipulating this internal tree-representation. The DOM tree is normally kept hidden from the user of the browser but can be made visible in every browser using development tools. An inspection of the DOM tree will tell web developers how the browser has interpreted the web page and help them resolve bugs.

### 2.1.2 JavaScript

Static web pages out of HTML and CSS are fine for displaying a simple website but soon web developers wanted to create more interactive pages. Initially they just were looking something that would reduce web traffic by taking care of simple tasks like form validation without having to would communicate with the backend server using the browser functionality. This wish was picked up by the developers behind the Netscape browser and LiveScript was born which was later renamed to JavaScript [21, 22].

JavaScript gave developers the ability to alter and interact with elements stored with a browser's DOM tree after the page was loaded. This concept proved to be popular and many other web browsers vendors started to incorporate JavaScript into their product. The rise of JavaScript was unfortunately in the middle of the so-called browser wars [23] when two of the major browser vendors, Netscape and Microsoft, were competing head to head adding ever more features to their respective browsers in order to gain the biggest share of the market. This battle in combination with the W3C not being able to get parties involved to agree upon a common set of web standards created an environment where a piece of JavaScript would work in one browser but not in the other. These incompatibilities due to the lack of standards let to such frustrations that the majority of software engineers turned against using JavaScript and adopted the practice of the backend server languages like Java [11] and PHP [14] to built web applications.

### 2.1.3 Ajax, HTTP Requests, JSON and XML

At the end of the browser war in 1998, Microsoft had won which resulted in a freeze in browser development. For about five years no new features were added which created a window of opportunity for not only other browsers like Mozilla Firefox and Opera but also many other web developers to catch up and innovate [24]. This renewal of interest in web technology lead to a overall embracement of web standards since all parties involved had realized that without open web standards like the W3C was promoting Microsoft would crush them all in new browser war. The growing popularity of Mozilla Firefox even meant more backend engineers started coding web pages with browser-native technologies, all attracted by the fact that they did not have to program a separate version for every web browser (except Internet Explorer). This trend is still going with browsers rapidly becoming fully standards compliant which already has lead to a sprint of innovative web applications coming onto the consumer market like Twitter [25].

A second trend in web development is to move from traditional desktop applications to web applications [10]. A key component in these applications is a technology called Ajax although we won't go into the exact details on the inner-workings of Ajax, since many others [26, 27, 28, 29] have already done that, we should explain briefly why it is currently being used in many websites.

Ajax stands for "Asynchronous JavaScript and XML" and it refers to the use of the XMLHttpRequest API (XHR) is to transfer JSON [29] or XML [30, 35] formatted data between client (the browser) and server. The XMLHttpRequest object [10] allows an asynchronously fetch data from the server out of sight of the end user. In this way one can update parts of a webpage without having to reload the entirely and thus limit the amount a lot of data-traffic needed to display and access information.

The beauty of Ajax technology is that it works seamlessly with HTML and with help of JavaScript rich graphical user interface (GUI) capabilities can be added without having to re-implement content. Furthermore Ajax is compatible via the Hypertext Transfer Protocol (HTTP) with almost all commonly used application server-platforms like PHP: Hypertext Preprocessor (PHP), Active Server Pages for .NET (ASP.NET) and the Java 2 Platform, Enterprise Edition (J2EE).

### 2.1.4 JavaScript Libraries: Prototype & Script.aculo.us

Programming browser-native technology cross-browser is still rather complex and cumbersome since web technology is moving at such a pace that standards are being agreed upon along the way after implementation. We also still have some legacy incompatibilities within the interpretation of current standards especially Microsoft's Internet Explorer has its quirks. The main reason for this is that historically Microsoft has a tendency to define it own web standards. The introduction of open-source JavaScript libraries around 2005 like Prototype, jQuery and Dojo [31, 32, 33] are all attempts to make developing with JavaScript cross browser easier for the average developer. Each library offers a set of routines to manipulate or walkthrough the Document Object Model of a webpage and usually the calling and handling of Ajax XMLHttpRequests has also been greatly simplified. This has meant that developers now have to spend less time fixing browser incompatibilities and can instantly start building on top of an almost cross browser unified web platform.

## 2.2 Browser-native visualization technologies

### 2.2.1 VML

In 1998 the W3C asked for submissions of an XML-based web vector graphics language leading Microsoft partnered with MacroMedia (now Adobe Systems Inc) to submit Vector Markup language (VML) [34, 35] to the W3C as the new standard for graphics on the web. VML was intended to be an extension on HTML rather than a complete separate entity which has the benefit that it can be easily integrated into a webpage. If one takes a look at the specifications of VML you will see that the syntax is using a subset of XML similar to HTML statements. In addition to supporting styling with CSS VML elements may also be rotated, flipped or grouped. VML however offers only a limited set of options when it comes to image manipulation, built-in functions can only change gain, blacklevel and gamma or greyscale the image. Internet Explorer is currently the only mainstream web browser that offers native support for VML, since the W3C never made it a web standard and it remained therefore a Microsoft proprietary standard.

### 2.2.2 SVG

Scalable Vector Graphics (SVG) [36, 37] was developed by W3C in 1999 based on a combination of all the web vector graphic languages that were submitted including VML and PGML [36]. Many concepts like predefined objects, the use of XML, manipulating and grouping of objects were all incorporated into SVG. In contrast to VML which is extending HTML, SVG is a completely difference subset language of XML. This has as an effect that integrating SVG code within HTML document is quite different than VML.

First one always has to define a SVG root DOM element to which all SVG elements are attached to whereas you can put a VML object anywhere in the body of a HTML page.
SVG is unlike VML a W3C web standard and hence supported by all mainstream browsers except Internet Explorer which needs a plug-in in order to be able to handle SVG content.

SVG was designed from the start to offer a replacement for bitmap graphics which have some serious downsides when used on the Web. Scaling of a GIF, JPG or PNG image will always reduce quality and for significant alteration of images a graphics software package is needed. SVG on the other hand is stored as plain text which can be edited with any text editor and due to its vector nature can be infinitely scaled. You do not have to edited the plain text file if you want alter a SVG image in fact there is a good range of graphic software packages that can help you to create and manipulate SVG-based graphics.

Another issue with bitmap graphics is no support for internationalization which is a problem if you for instance want your corporate logo to have a subtext in the customer's language. SVG can resolve this issue by reading out the customer's system language setting and change the graphics to the desired language. Other features included in the SVG language are animated and interactive graphics moreover a special subset named SVG Tiny has been created for mobile phones. The possibilities of SVG seems endless yet widespread adoption of the technology is still halted since Microsoft's Internet Explorer which currently holds the biggest market share does not support SVG out-of-the-box.

Looking at how SVG can be applied for manipulating images within the browser then we have to incorporate SVG filters. SVG Filters can be used to apply almost any effect to an image by specifying your own mathematical formulas to change an image colours or by using one of the built-in filter functions.

2.2.3   Canvas

Canvas [38] was conceived by Apple Inc., in 2005 to make development of desktop widgets for their operating system Mac OS easier. Whilst SVG and VML are both initiatives from 1998 and 1999, Canvas is a lot younger however after some initial criticism it has become part of the upcoming HTML 5 standard.

Critics argued that there was no need for yet another web graphics standard since SVG was already sufficient for the job at hand. Furthermore Canvas by design does not offer a way to save your graphics into a descriptive language like XML it only offers the option to export to one of the several bitmap graphics formats.

The name 'Canvas' basically depicts what the technology does; after definition of a <canvas> element in the DOM tree then one can paint anything on it using JavaScript. Most developers find Canvas is easy to learn and it has no notion of XML thus saving the normal overhead of XML which we normally needed in SVG to render graphics. The only drawback of Canvas is that drawn elements are not visible in the DOM tree which makes it harder to understand how a webpage is working by looking at its source code. Nonetheless almost all major browser vendors expect Microsoft support and have implemented Canvas in their products since it is seen by most as a future alternative solution to using Adobe Flash.

# 3. Creation of a prototype

We could have programmed a simple proof of concept with a couple HTML files to research the subject of this paper. However we wanted to get the complete picture of what it takes to design and implement a web application that used browser-native technologies for image manipulation. Our aim is to develop a fully functional application demonstrating what is possible with browser-native technology. Although the application will not be ready for commercial deployment straightaway it should be a working prototype with a couple of remarks on how to use it.

## 3.1 The beginning

### 3.1.1 The overall idea and design

The whole idea behind this paper is to see what possible with the current state of web technology and what will work cross-browser. It also an attempt to break free from the traditional way of how photos management and sharing is currently done by several of the high profile websites. Most of these like Flickr, PhotoBucket and Picasa use a grid-like layout to display collections of images and do only offer a very basic set of image operations. We would like to try to change this by offering users an application in which they can organize their photos like there are accustomed to with a real-life scrapbook.

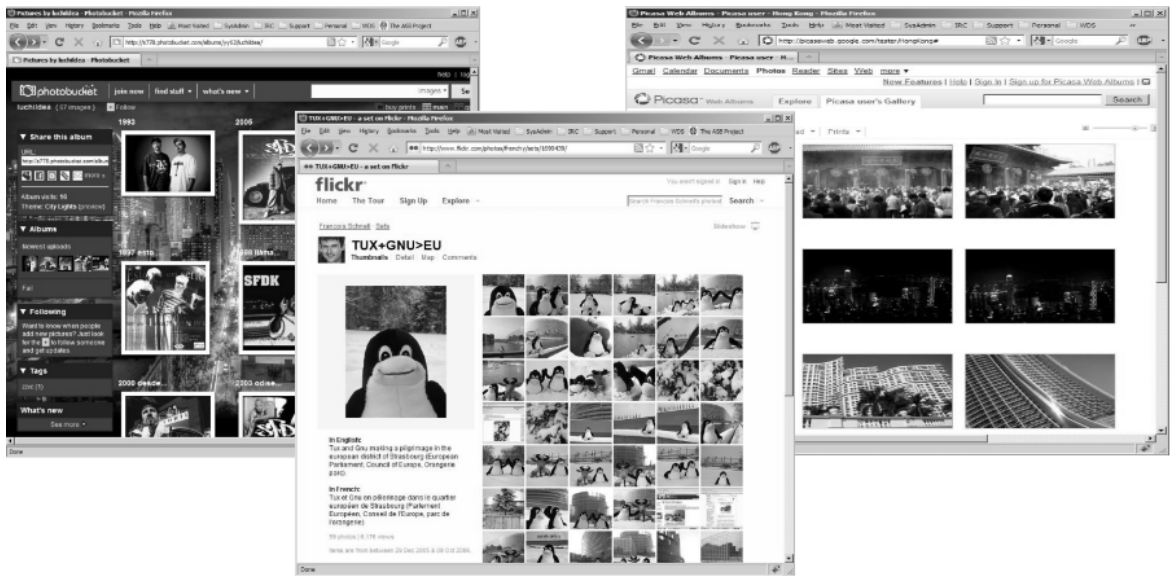

**Figure 1**. Impression of Flickr, PhotoBucket and Picasa websites showing the grid-like layout

This will be our starting point, something as simple a photo scrapbook in which users can position his photos however he likes it. Our aim is to give the users of our application the freedom back that they lost when the world moved from the analog to digital photography. This move and the rise of the internet changed the way how most of us share photographed experiences. A vast majority of the people who are connected to the Internet now upload their photos to websites like Flickr or Photobucket to share them with friends, family and the rest of the world.

The World Wide Web and its underlying standards however have put several restrictions on how information can be displayed for instance everything on the web needs to be square. This may sound strange to a person who never has built a website but this is the daily reality of a web developer. Round corners or fluid designs are usually created by a square image in combination with either transparency or the background color of the image equals the background color of the website.

The technology behind the Internet is however rapidly evolving and engineers have added standards that allow web designers and developers to think out of the square box.
Web graphics technology makes it possible to use the web as a proper canvas and to draw any shape or size you can imagine. No need to use small images to make the round corners but just a few lines of code. The sad thing is that several of the standards and languages who make this possible have been around for more than a decade. They are however still hardly used since they are either unknown to or considered as too complicated by most programmers. We are now taking another look at web graphics technology to see if we can mix them up with Ajax and use them to create something special.

Mimicking a regular photo scrapbook has the implication that we should support quite common actions as moving and rotating a photo. This means a photo should be positionable anywhere within a webpage and even overlapping of photos should be possible. The ability to stack photos slightly on top of each other might not be as efficient as the grid but it does create a more natural look and feel for the end user.

We should also incorporate the benefits of the digital era and incorporate cropping, scaling and maybe some effects if possible. Effects like grayscale, changing brightness, contrast and playing around with the color are quite handy in either fixing a poor taken photograph or in the case when adding an effect just gives it the little bit of magic needed to create a prefect picture. Another feature that would be great to have is the ability to write text on your picture as one would with a normal photo. We are thinking of a digital version of the good old polaroids on which you could handwrite a text on the bottom of every picture.

Our vision has been set out and now the time has come to see what many lines of code and hundreds of working hours can accomplish. To summarize our prototype should include the following features:

- 360 rotation, cropping, scaling and stacking of photos
- Ability to drag and drop photos anywhere within the webpage
- Some basics image manipulation operations including brightness, contrast, grayscale
- Adding borders and text to photos: creating polaroids and perhaps support the function to handwrite on a photos.
- All functionality should be implemented to work cross-browser so it's accessible by the majority of all web browsers.

The entire list above is of course subject to the fact whether we can implement a feature with existing technology and within the deadlines. We will discuss the list again at the end of this chapter to evaluate what we did and did not get working and the reasons behind its success or failure.

### 3.1.2 Facing the first challenges

We are building a prototype web application and by doing so there are several challenges that we could think of even before one line of code was written. The aforementioned browser incompatibilities with regards to how to implement web standards is off course a major hurdle but there are multiple issues that came to mind when we were brainstorming on how to implement the prototype.

#### 3.1.2.1 Different screen sizes

The first challenge is simple in nature there are multiple computer screen sizes around the world ranging from 3.5 inch screen in an Apple IPhone to a 42 inch LCD TV attached to a media center. Taking into account that we want build a web interface were users can place anywhere within the browser window then we need to figure out how to preserve the same experience for the users across different screen sizes. Although at moment in time we not planning support for devices like the IPhone still people will access and use the application using different browsers on a variety of devices and machines.

This raises the question how to handle a photo that is 400 pixels to the left and 800 pixels from the top on a 30 inch screen, which is 2560 by 1600 pixels, and displaying it correctly on a netbook with mere 1366 by 768 pixels resolution. The challenge that lies ahead is to come up with a mechanism that will use a combination of scaling and moving photos in order to preserve the same user experience across different browser window sizes.

#### 3.1.2.2 Human interaction

We were also contemplating on how users were going to interact with our web application. It's a fact that all computer users will have a mouse and a keyboard however we immediately identified a problem: Apple Macs. Macs are not only beautifully designed but they also follow a different graphical interface design paradigm than Windows or Linux.

One of the quirks of Apple operating system is that until recently everything was designed for use with one single mouse button. As a result Apple's mice are all one buttoned mice and if a right mouse click is needed one has to hold down the CTRL-key. The more recent models of Apple solve the one button issue by using a multi-touch trackpad but the overall majority of the Mac users will be restricted to one button.

A similar mouse issue occurs within every version of the Opera browser which by default is configured not to allow JavaScript to receive right click mouse events. On a right mouse click within Opera the browser's context menu will always appear making it impossible to attach our own menus on a right click. One of our goals is to be cross-browser and preferably we want the same universal way of interaction in all of them. This is one of those problems that sounds so simple but if we do not get it right than we will cripple the usability of our web application.

## 3.2 Implementation

### 3.2.1 The blueprint

After creation of the task list a raw design of our application was needed so we could visualize our ideas in order to make up a planning. Figure 2 shows a mock-up of the initial page a user should get after he or she has logged into the web application.

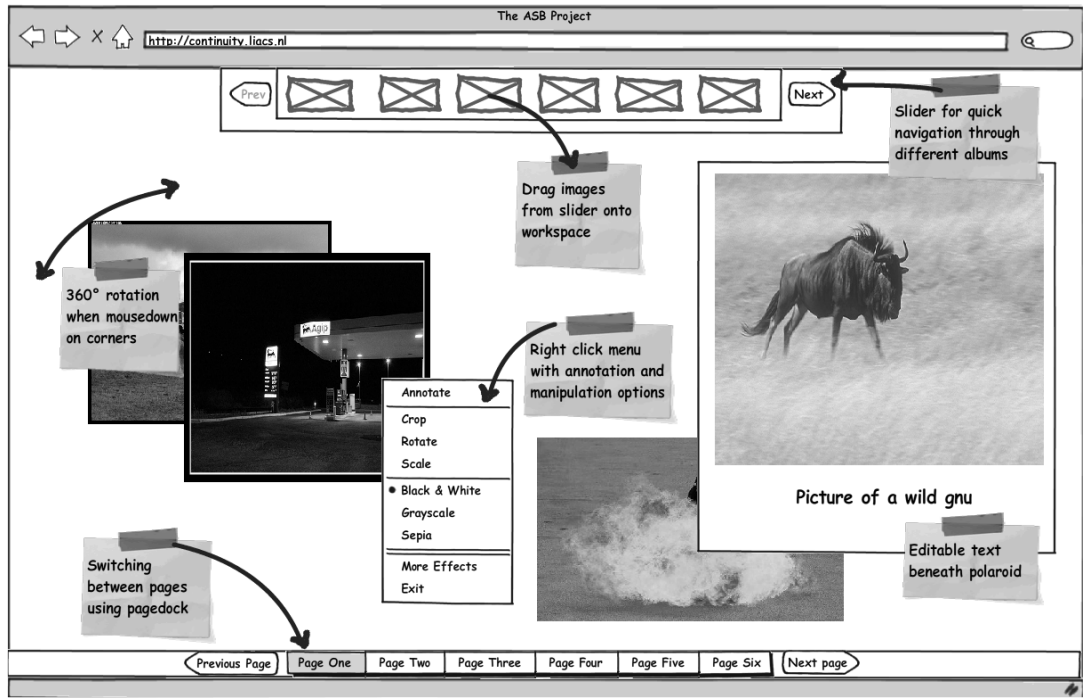

**Figure 2.** Mock-up of the main web page web describing the overall interface

Now we have designed the look the next step is to write up the underlying architecture with details on how everything is interconnected. We decided to built upon the well-known Model-View-Controller architectural pattern and figure 3 demonstrates how we applied it to our application

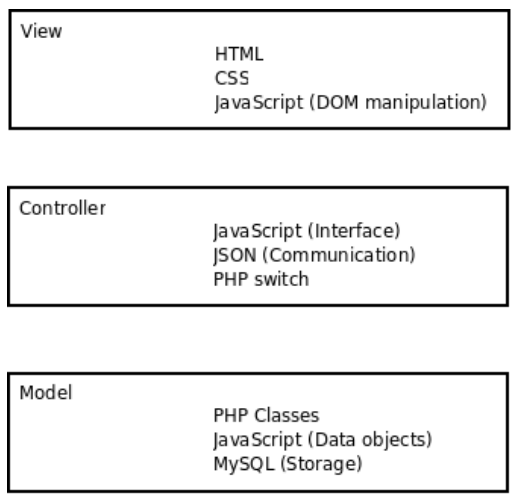

**Figure 3.** The MVC architecture as used with the prototype

Taking the MVC model as a guideline we implemented PHP classes and JavaScript objects into the three categories; visualization, logic and the model. On load of the main HTML/PHP file by the browser the execution of the main JavaScript file is triggered. This central piece of JavaScript will then initiate the rendering of three other JavaScript objects: the scroller, pagedock and the workspace. The entire application basically consist out of one small main HTML file and several JavaScript files. Everything needed to create the application is dynamically created by JavaScript attaching elements (styled with CSS) to the DOM tree.

The main JavaScript file also has controller logic built in that processes and responds to the users input. Furthermore it contains an Ajax handling layer which is providing other JS objects with a way to communicate via JSON statements with the backend 'PHP switch'. This switch interprets the JSON statements and calls the needed PHP objects to process for instance a user authentication request. One unique aspect our program is that we have opted to implement the several objects within our model as well in PHP classes as in JavaScript objects. This has the added benefit that whilst developing the frontend you only have to think in objects and their methods instead of having to manually call a particular backend function. This object oriented notion in the frontend also almost eliminates the need to think about communication with the backend. For example if you want get the description of a photo but it is currently not in the JavaScript photo object then the object itself will on request call the backend, get the missing piece data and return it to its requester.

Figure 4 displays the model we used within the application to handle and store all data into the database. The relations between the objects are depicted with one-one or one-to-many relations and act as restrictions defining the set of valid actions by the end user. The faces and templates objects might seem weird but they are a legacy component from the time that also face recognition was on our task list. We made the decision to leave these two components in the application since face recognition will probably be a desired future feature after initial release of the application.

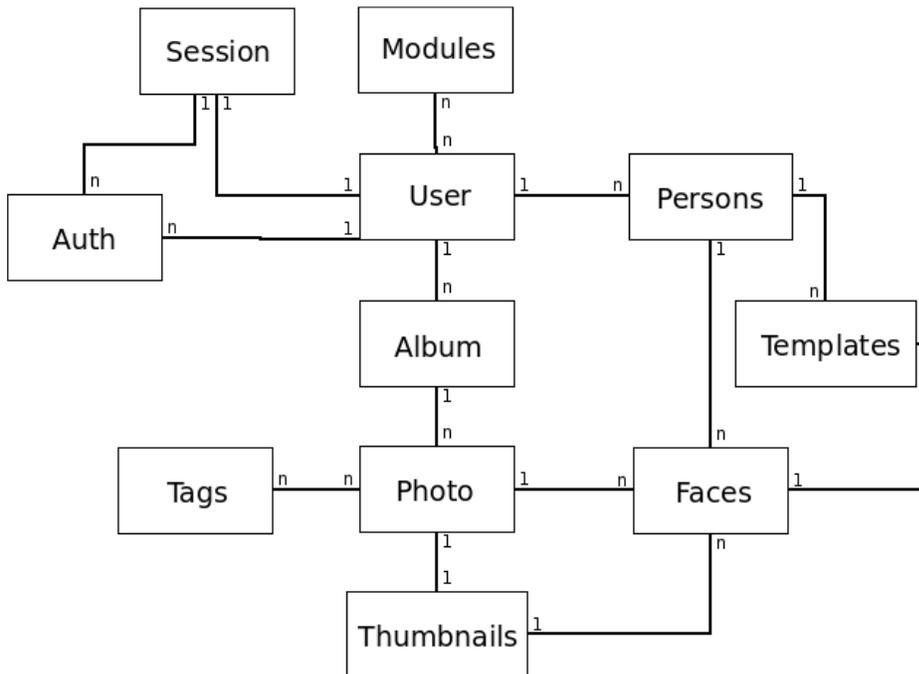

**Figure 4.** A UML model of the relations between the data objects used within the application

3.2.2  Developing the frontend

The frontend user interface that is running in the client browser consist as we already mentioned for the most part out of JavaScript and in order to get the same usable application cross-browser we needed to comply with interpretations by browser vendors of current web standards. We started development off by looking at ways to make implementation easier and faster since we only had a certain timeframe in which we needed to deliver a working prototype.

3.2.2.1  Choosing a library

The usage of one of the many JavaScript libraries within the frontend was an obvious choice since they level the playing field by providing several needed DOM traversal and manipulation functions that just work regarding the browser used.  We needed one that was relative small in size, extended JavaScript with basic data types and functions that are normally not found in the language and that was well-supported by a big community of users.

The reasons behind these three demands are simple if a library is too bloated in size the download to the user with a limited bandwidth is to big resulting in loading time of our application hitting more than several seconds.  Extending JavaScript is needed since set out the idea to have the same data structure in the frontend as in the backend. In order to realize this we need JavaScript to be able to have certain data types and have several functions that are standard in PHP but not in JavaScript. Development goes also faster if you are accustomed to programming in a certain way like in PHP especially if one needs to solve complicated design problems.

At the beginning stage of the development of our frontend there were three candidate libraries that met our criteria: Mootools, Dojo and the Prototype & Script.aculo.us combination [31, 33, 39]. This was back in 2006 since that time several other JavaScript frameworks have been released with all of them having their own unique capabilities and feature. There has always been a big difference between libraries in what has implemented so we did a simple comparison by searching the web on what the overall JavaScript community was doing and who was making the most progress after which we settled on using the Prototype & Script.aculo.us combination. The main reason for this choice was steady release of new versions, small file size and it still is one of the popular frameworks around.

3.2.2.2  One block at a time

We always knew that even with a JavaScript library what we were working on was not going to be a walk in the park therefore we needed a way to keep it manageable. By chunking up the entire project into smaller object-oriented we should be able to design and test concepts faster.

The first thing on our list was to get a drag & drop rotatable photo working in every browser. It quickly became apparent that basic HTML mixed with vector graphics was insufficient and just was not working for us. Vector graphics technology could get the job done but the traditional examples of how to do things were incompatible with our feature list. If implemented by the book the backend server had to detect which browser was making the request and based upon that detection sent the right set vector graphics files either SVG or VML.

This approach would have meant implementing everything three times which is not a feasible plan. Luckily however we already had started to work out the basic graphical user interface on which the vector graphics photos would be rendered. It already included a photo scroller, workspace and page navigation bar and it was completely built using nothing but JavaScript. This gave us the idea to do the same with SVG and VML and just use JavaScript to build up the DOM tree in exactly the way as the traditional file based approach would have do it.

3.2.2.3   Abstract objects

The use of JavaScript to do web graphics in Canvas, SVG and VML is not new however most developers only apply JavaScript to change object after they have been rendered by a vector graphics HTML file. There is a significant programming challenge if you want to render everything with JavaScript and quick web search revealed two libraries already existed that could do the job: Dojo and Raphael [33, 40].

A choice for Dojo which supports all web graphics technologies would have meant ditching the Prototype & Script.aculo.us combination, learn Dojo and rewrite all the code that we had written up to that point. This was not an option since we simply did not have the time so Rapheal was our only option with the added drawback that it did not support Canvas. The easiest solution would be to scrap Canvas as a way to do image manipulation on the web yet in our preparations we had already learned that it was the most capable language to do the job.
An second option was to extend Raphael with Canvas support and then build our application on top of the newly created library. A quick look at the code behind Raphael showed us its inner workings and we made the decision that it would probably be easier to rewrite everything ourselves. By doing our own implementation we would have full control giving us the ability to only implement the things that we needed making overall code base smaller and more flexible.

The next challenge was how to combine three different web graphics languages into a single JavaScript fill that would fulfil our feature list. Again we had a look at how Raphael was implemented which gave us the solution by creating a separate abstract photo data object from which we would be able to generate photos in the three languages (see figure 5).

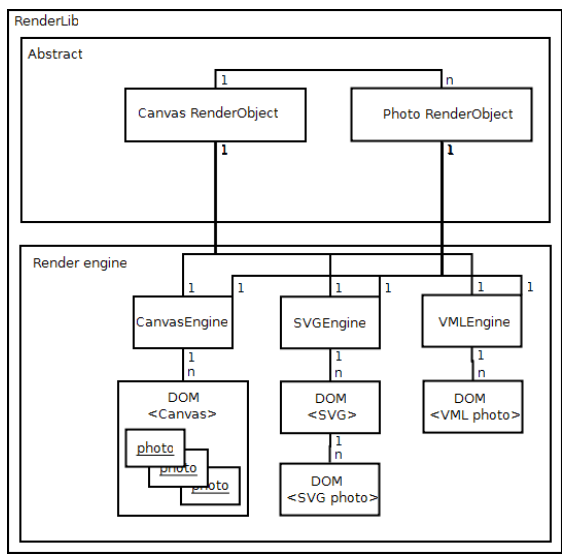

**Figure 5.** Representation of the internals of our JavaScript web graphics rendering engine

The abstract Photo RenderObject holds all the regular attributes of the photo such as width, height, source URL of the image and position of the image relative to the canvas and so on. Image operations such as rotate, scale and drag & drop are first calculated with the abstract and the resulting position, size and angle is than passed on the respective Canvas, SVG and VML rendering engine.

The differences in the inner workings of the several web graphics languages were the cause to develop to individual rendering photo strategies for each one. In order to create a drawing with Canvas one first need to create a <canvas> element, attach it to the DOM tree and then use JavaScript to draw upon this element. Photos drawn with Canvas are inner manipulations of the <canvas> element and therefore they are neither attached nor visible in the DOM tree.
Canvas literally acts as a painting canvas meaning you draw every pixels of image and when for instance move an image you have to wipe the canvas clean and draw the image again on the new position. SVG and VML are completely the opposite of Canvas namely on creation of a photo a new element will be attached to the DOM tree. One of the differences between SVG and VML is that SVG has to have a root <svg> element in the DOM tree to which all SVG content is attached whereas VML elements can, after defining its namespace, be inserted at any point in the DOM tree. The abstract class Canvas RenderObject eliminates all differences in initial setup and depending on language creates a <canvas> or <svg> element or defines the VML namespace.

3.2.2.4   VML

Programming VML on itself is quite tricky since the language is not actively being used by a lot of the mainstream web developers making code examples scarce. Microsoft, the vender who is backing the language, has not put any serious development effort into the language after it won the browser war with Netscape. This lack of its creator's attention is noticeable in VML capabilities to manipulate images within Internet Explorer.  The language is full with its own quirks of doing certain operations that is out of touch with the modern practices. A typical example of a quirk can be seen when cropping an image; this operation can only be done using percentages instead of the more common use of pixels. A second one can be found in the application of image effects which all have been implemented in VML using separate DOM attributes. Although this makes applying an effect easy nowadays no developer would design a language in this way since it would mean creating a DOM attribute for every supported effect. Effects like black & white, greyscale are just booleans within every VML image element whilst black level, gain and gamma can be set by changing the attribute's value. The previous line pretty much sums up which effects are by default available in VML, a couple more can be gained by calling Internet Explorer's *BasicImage* filter namely blur, inverting colours, opacity, mirroring and X-ray.

This give us a very limited set in order to recreate the basic effects as brightness, hue and saturation and we have tried to combine multiple combinations of the available functions but unfortunately we were unable recreate all of them. In the case of brightness and contrast we were able to get close to but still not near enough to how the effect looks like in SVG or Canvas.
The image manipulation limitations of VML are mainly the result of it's the proprietary nature and poor documentation which does not allows us to see how effects have been implemented by the IE development team. We were unable to find any recent press releases by Microsoft on VML's future development which let us to believe that VML is basically a dead technology that is kept alive by Microsoft for the sake of backwards compatibility and obligations to third parties.

### 3.2.2.5 SVG

In contrast to VML, SVG was the easiest technology to implement into the rendering engine since it is a W3C standard and therefore has been well documented. The only trouble we had was founding out how to recreate inline SVG statements from a JavaScript file. It took us about half a day to figure out how to properly code it since this is something that is not described in the official documentation. We discovered whilst working with SVG that it is quite a powerful language although harnessing that power is something that requires a lot of time and skills. Once we figured out how to use SVG we used it as a stepping stone to debug the other two web graphics languages.

Looking at image manipulation there is a wide range of SVG filters to accomplish a wide variety of effects. There is a downside however that all effects have to be specified in a mathematical formula form or exist in the set of SVG filter functions. Effects like greyscale, sepia, brightness and hue can easily be created but something simply as turning the image into black and white values is more cumbersome if not almost impossible.

### 3.2.2.6 Canvas

Canvas has the strongest papers for image manipulation since it has the capability to access and change a single pixel on the drawing canvas using a few lines of JavaScript code. Any existing image effect can therefore be recreated within the browser; the only limiting factor to the technology is the JavaScript processing speed of the web browser. The more pixels manipulated, the more CPU and memory will logically be consummated and the time before a particular image effect is shown to the user will be longer. This explains why the focus in the currently ongoing browser war between Firefox, Opera and Chrome primarily focuses on making the processing of JavaScript significantly faster with every release. The bigger and more complex web applications will become in the future the better the JavaScript engine of the browser needs to be in order to make an application usable for the end user. As part of the upcoming HTML 5 standard, Canvas has a huge potential and therefore although still in its draft phase has already been implemented by all the major browsers except Internet Explorer.

Back to our prototype implementing the basics like moving, rotate and scaling were relatively easy to get working. The Canvas render engine feed of the abstract photo model and draws the given photo object onto the canvas embedded within the DOM tree.  This all goes fine for a rendering a single photo however when switching to multiple photos several issues emerged. The first thing that needed to be tackled was something simple as z-indexes which are used within the DOM tree to determine whether an element is to be rendered on top or behind other elements in the tree.

There are no DOM elements within < canvas> element  it's just a blank canvas consequently if we want to render multiple images and be able to change which one was in front of the other then we need to devise our own mechanism to do this. The solution came in implementing a global array within the abstract Canvas Render Object that stores that the order of photos on the z-axis. In addition we programmed an algorithm that would update the z-indexes of all photo objects affected when a z-axis change was made to one of the images in the array. Whenever a photo object is added to the array it gets a z-index based on the number of photos already in the array plus the z-index of the <canvas> element upon which we are drawing. The number of different values for z-indexes is thereby always equal to the number of photos in the array.

On a down-shift, e.g. top element A is send to the back; we eject A from the array then perform a right shift on all photos in the array from the point where point A needs to be inserted till the point where A originally was located. During the shifting process we adjust the z-indexes of the photos according and as a final step A is again inserted into the array on the empty slot of the z-index were the user wants the photo to be placed. The 'bring to the front' algorithm works similar only we then use a left shift to re-arrange the photos to the z-index change.

The z-axis problem was now resolved but on every move, scale or rotation of a photo meant that the entire <canvas> element was being redrawn and all photos on it. This resulted in a very flickering webpage since all photos behind the photo being dragged flickered or disappeared until completion of the operation. JavaScript was, in the majority of the Canvas capable browsers, just not fast enough to render everything smoothly. In order to keep the user's experience similar to SVG and VML we decided to implement two rendering layers.

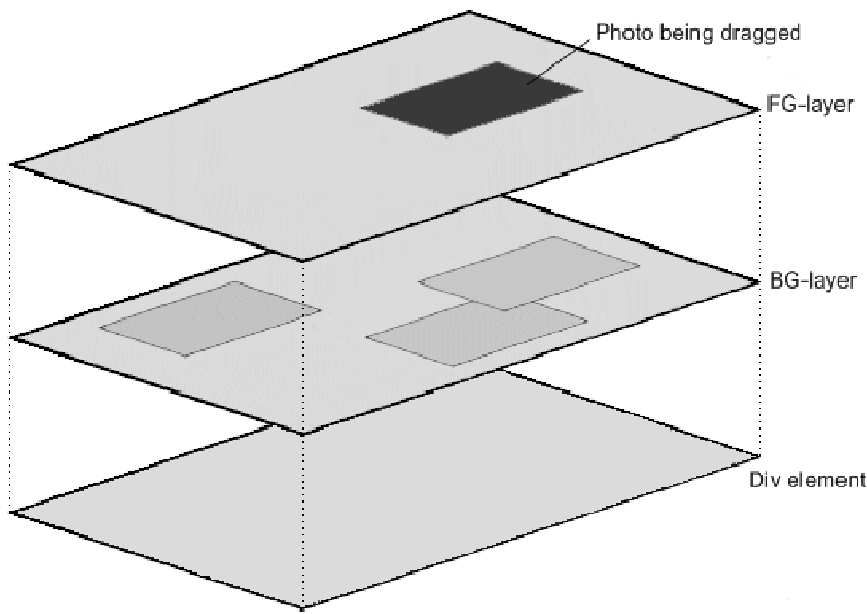

**Figure 6.** Representation of the Canvas layer rendering system

On creation of the JavaScript rendering engine two <canvas> elements are attached to the DOM tree (in our application also bottom Div element is inserted which is holding background picture). The BG-layer <canvas> element holds all static content whilst the FG-layer is where interactive content are is drawn upon. This separation eliminates the stuttering of the photos in the background and also saves CPU time since only dynamic content will be redrawn during the length of an interaction.

The third issue that we encountered had to do with the fact that Canvas is not yet a full W3C web standard. Although several browser vendors have already implemented large chunks of the standard still several pieces are still not implemented. One of the unsupported Canvas features is the ability to write text onto the Canvas. We ran into this fact when we wanted to implement writing text on top of a photo. At the moment of writing this paper in August 2009 only Apple's Safari and Mozilla Firefox support the *strokeText* function and we therefore had to resort to implementing our own text function in order to write on the <canvas> element.

3.2.2.7  Encountered problems during development

Trying built a web application that for the most part consists out of JavaScript is something not most web developers will likely do. Most of the web development community is primarily focus on building corporate web pages not web applications. JavaScript is primarily used to add little interactive touches to a webpage and developing a large scale web application with it is something only small percentage are doing.  There is also not one vendor standing behind the language that provides you with ready to go guide with coding examples of how to do everything. This has as a result that developing a web application with primarily JavaScript is something of a quest through a large variety of websites in order to find out why something little like knowing when an image is completely loaded does not work in a particular browser.

One has to have a good understanding of the inner workings of JavaScript and its prototyping nature before one is capable of debugging its errors. How to debug large blocks of JavaScript code is a skill we had to learn since as a scripting language there is no compiler to show you at which line you made a mistake. Firebug [41], a well-known Firefox debugger, can help quite significantly but it frequently happens that the line with the bug is a dozen lines away of where it was reported.  Debuggers for Opera and Internet Explorer do give you a insight into the DOM tree but are not well suited for debugging JavaScript. In the case of Internet Explorer a proper debugger is on the wish list of many web developers since the browser has the habit of crashing if it is not capable to process a particular sequence of JavaScript code.
JavaScript development is therefore one step forward, two steps back in order to make sure previous developed code works in every browser. A good code versioning system like CVS or Subversion [42] helped us to jump back in time to a working version and start implementation of a particular feature again after with had come into a debugging deadlock.

Debugging during execution of the code is possible using Firebug's watch statements and breakpoints but we found out that implementing PHP error code system in JavaScript is far more time efficient. Although almost never used by the majority of web developers JavaScript does offer the notion of try-catch error handling. The only issue it has is that unlike PHP you cannot attach error codes to an error event but using JSON we were able to add this functionality to JavaScript. Our application will now give you a human readable message with the error code on runtime error which allows you to look up the cause in our documentation. A secondary benefit of the error handling scheme is that an error in the backend can be processed by the frontend since both support the same error data type.  The downside of the system is the mandatory implementation of validation checks in every object and classes verifying the correct working. This overhead in front and backend will make the overall execution of the web application slower but we have found it does not outweigh the benefits of being able to fix runtime errors.

3.2.2.8  Resolving open issues

In paragraph 3.1.2 we discussed two challenges that we were facing at the start of developing the prototype; different screen sizes and human interaction. The screen size issue can actually be relatively easily resolved using web graphics technology. Canvas offers the option to change the scale of the rendering canvas giving you to option to scale all to be render objects at once. In SVG & VML a similar effect can be accomplished by nesting all rendered object into a *group* element which is then scaled relative to the screen of the end user.

A simple algorithm to resolve the screen issue would be to save all object's angle, position and size to a standard size browser window say 1024 by 768 pixels. On request by a user with a different screen size than standard the object's attributes are scaled to overcome the difference.

The second challenge was how to cope with different mouse interfaces of the different browsers which proved to be a more complex problem. Implementing a right mouse click cross-browser is just not feasible and the only solution we found to the problem is to use a mix of left mouse click and a key press. The use of web graphics even expanded the problem since even in Firefox it is difficult to attach a custom routine to a right mouse click event when this event is triggered over an image. Our solution is for a usability point of view not the best solution but it does provided a universal interface independent of the type of browser one uses.

### 3.3   Backend services

#### 3.3.1   MySQL and PHP

The backend framework of our web application is fairly straightforward consisting out of several PHP files and a MySQL database. One of the unique aspects of our backend is that we programmed it in such a way that all communications with the outside world go via a single main PHP file. This main file responds when visiting the web address by generating a HTML file which on load by the browser dynamically builds up a user interface using JavaScript. From that point on all communications with the backend server are done via Ajax in combination with JSON giving the user an experience similar to a desktop application. By tunnelling all traffic through a single PHP file it is easier to debug and secure the web application.

A second feature that we implemented was a failover in case a web graphics technology does not support a particular image manipulation operation. VML had such a limited set of operations that we decided to incorporate a failover to traditional server based processing. This backup mechanism ensures that all features will always work for the user in every browser although execution time of an image manipulation operation may rise if a roundtrip to the backend server is needed. The server based processing of images was easily implemented by the use of a third party PHP library that incorporates GD library which is PHP de facto image manipulation package.

### 3.4   The final prototype

#### 3.4.1   The check list

On our task list were the following features that should be incorporated to answer the question whether browser-native technology can be used to build an image manipulation web application.

- 360 rotation, cropping and scaling of photos
- Ability to drag and drop photos anywhere within the webpage
- Some basics image manipulation operations including brightness, contrast, grayscale
- Adding borders and text to photos: creating polaroids and perhaps support the function to handwrite on a photos.
- All functionality should be implemented to work cross-browser so it's accessible by the majority of all web browsers.

In the following paragraphs we discuss several of the items on the list and explain how successful we were in implementing them or what issue still need to be resolved.

3.4.2   Comparison on image manipulation operations

Table 1 shows which image manipulation operations can be implemented using Canvas, SVG or VML. Our prototype incorporates all the manipulations that are possible in all three languages plus sepia for Canvas and SVG. As we discussed earlier Canvas is the most powerful technology when it comes to images manipulation although we have to mention that during development we have noticed that there are limits to the number of photos that can be displayed, we will investigate this further in the next chapter.

| Description | Canvas | SVG | VML | Remarks |
| --- | --- | --- | --- | --- |
| Black & white | ✓ | ✓ | ✓ | SVG not 100% perfect, still gray values visible |
| Blur | ✓ | ✓ | ✓ | VML using IE BasicImage filter |
| Brightness | ✓ | ✓ | ✓ | Using VML's black level but effect not similar to Canvas & SVG |
| Border | ✓ | ✓ | ✓ | |
| Contrast | ✓ | ✓ | ✓ | VML using gain function but effect not similar to Canvas & SVG |
| Cropping | ✓ | ✓ | ✓ | |
| Desaturate | ✓ | ✓ | ✓ | VML using IE BasicImage filter |
| Drag & drop | ✓ | ✓ | ✓ | |
| Emboss | ✓ | ✗ | ✓ | |
| Flip horizontal | ✓ | ? | ✓ | SVG: maybe possible needs further research, VML using IE BasicImage filter |
| Flip vertical | ✓ | ? | ✓ | SVG: maybe possible needs further research, VML using IE BasicImage filter |
| Greyscale | ✓ | ✓ | ✓ | |
| Hue | ✓ | ✓ | ✗ | |
| Invert colours | ✓ | ✓ | ✓ | |
| Opacity | ✓ | ✓ | ✓ | |
| Red eye removal | ✓ | ✗ | ✗ | |
| Rotation 360° | ✓ | ✓ | ✓ | |
| Saturation | ✓ | ✓ | ✗ | |
| Sepia | ✓ | ✓ | ✗ | |
| Sharpen | ✓ | ✓ | ✗ | |

**Table 1.** Comparison of web graphics on image manipulation operations

Implementation of the various effects in one of the three languages was not that difficult; we had more trouble finding out the inner workings of an effect e.g. how it changed the RGB values of the image. As mentioned earlier, VML has almost no options to do image manipulation that affect the color of an image. During development we also found out that finding the right combination of SVG filters to accomplish an image effect can be quite challenging.

3.4.3   Text and handwriting

In SVG & VML there are two ways of creating texts either use fonts or use vector shapes to recreate fonts. The easiest solution would be off course to use fonts but if chosen you will be restricted to the ten default fonts that are available in every web browser. Using shapes one is able to render any desired font although one has to manually convert existing system fonts in their vector counterpart. Furthermore rendering of fonts with shapes is known to be a lot slower than to use the built in fonts. As explained in paragraph 3.2.2.6 Canvas will eventually support fonts but for now we had to resort to using shapes to draw fonts. After a lot of effort on our part we were able to get shapes working in Canvas and VML in order to create the sans font type due to limited time SVG still needs to be implemented.

We researched whether it was possible to support handwriting on the photos and we were able to find several examples on the web that proved handwriting is possible using web graphics. Looking at the source code we saw implementation is relatively easy; one takes the input mouse coordinates and every movement is translated into window coordinates which are then filled using a vector shape or line using Canvas, SVG or VML.  We have not included this feature in the web application yet since integration with web graphics render engine proved to be a bit cumbersome and time consuming hence we were forced to scrap it out  of the release prototype.

3.4.4   Developers perspective on using web graphics in a web application

At this point in the thesis we would like to highlight some aspects of developing a web application with web graphics that are important to developers when having to make a choice between the different web graphics technologies. In real-life there are two options either use Canvas & VML or SVG & VML, VML is always in the mix since Internet Explorer does not natively support Canvas or SVG. If we were developing commercially with a launch frame of 3 to 6 months we would choose the SVG VML combination since both languages are similar in design making developing rather easy. Both of them are well supported and have fast execution times making them ideal for highly interactive applications. Using Canvas and VML in a single web application is just not logical since it would require you to write separate rendering engines for each one, forcing you to implement every desired feature twice.

Looking at the image manipulation capabilities of Canvas in table 1 we sure it is going to be a technology that will be used a lot in future web applications. However at this moment in time Canvas is still not fully supported by most browsers and should therefore probably only be used on a small scale within a web application. In the near future, say beginning of 2011, we hope Microsoft has come to its senses, drops VML and starts supporting Canvas or SVG.  Developing for Internet Explorer and especially with VML is no walk in the park since the IE render engine is not very forgiving and at the tiniest mistake in your VML declaration none of the VML objects will be displayed.  There are already several web applications that use web graphics in displaying their content for instance the direction arrows in Google maps are drawn with SVG and VML. We believe Canvas and SVG both have a future and probably a mix of the two technologies will be used in the creation of new innovative web applications.

We have, before this prototype, developed several backend infrastructures for Flash and Java based web applications and we have to say compared to the one in this prototype there is not really a big difference between the technologies. All modern web applications use Ajax so most developers should familiar with developing a JSON processing backend.

If a developer chooses, like we have done, to go browser-native instead of using Flash then he will face some common roadblocks. Browser-native means adapting to the browser functionality and there are still differences in what each browser supports. Furthermore the native set of DOM traversal and manipulation functions is rather limited. The solution to the problem as we have done is to use a JavaScript library which eliminates the gaps and provides you with the functionality that should have been in JavaScript by design. We are actually of the opinion that every DOM traversal operation that is in the Prototype library should be incorporate into the next iteration of the JavaScript language since the majority of them are no more than logical. As we found out the hard way the use of a library also creates a dependency which makes it more difficult to jump ship if you needed a particular piece of functionality is in another library. Looking back we might have made the wrong choice in using Prototype since Dojo and jQuery libraries are currently more suited for developing web applications than Prototype. A JavaScript library does not solve all your problems there custom components in your application. In the development of these components you will encounter things that sound so simple but as we have learned the simplest things are usually the hardest to implement using browser-native technology. In our prototype for instance we would like to know when an image is completely loaded and there is a DOM Boolean attribute that will tell you whether this is the case. Still it took us quite some time before we had implemented this feature in such a way that it worked cross-browser.

As we mentioned in 3.2.2.7 developing a web application largely written in JavaScript is not something every web developer can easily master in a short period of time. Once you get the hang you of it you will be harvesting the rewards of building on open standards and your web application will no longer be dependent on a proprietary closed standard like Adobe Flash. Since all road paths of W3C standard technologies like Canvas and SVG are published on the web you will always be able to know what is going to happen in the near future giving you the ability to anticipate early on changes and new technologies. An additional advantage of building on open standards will be that your web application is likely to be more future proof to new developments like browsing on mobile devices.

To end this paragraph a couple of pointers for web developers:

- Choice your JavaScript library wisely although they do not explicitly mention it some focuses on building web sites others on web applications.
- Learn how to program Object-oriented JavaScript this will show you how, although of its prototype nature, most of the constructs available in PHP and Java can be replicated in JavaScript.
- Debugging large chunks of JavaScript is difficult but the pain can be reduced with the use of Firebug, version control and a good coding style.
- Canvas is currently the most talked about web language however it is not yet ready for large-scale production deployment; SVG and VML are better supported and execute faster than Canvas.
- Using SVG and VML it is best to first develop in SVG and than port the code to VML. The reason behind this is that it is easier to develop SVG with the help of Firebug than with the built-in debugger of Internet Explorer.

# 4. Experimental Results

The main research question of this thesis is to find out why browser-native technologies are not used to create image manipulation web applications whilst they have been around for quite some time. Our prototype has demonstrated that Canvas, SVG and VML are quite capable to perform image manipulation operations. In this chapter we are going to test the web graphics render engine in our prototype to see whether it will hold up in a series of experiments that simulate real world usage. We hope that the results of these experiments will give us an insight into any potential practical roadblocks that are preventing widespread usage of browser-native technology to perform image manipulation tasks.

## 4.1 Experimental setup

### 4.1.1 Hardware and infrastructure

The performance of any web application is directly linked to which browser you are using and its underlying operating system and hardware. We have chosen to do all our experiments on an Apple MacBook (Core 2 Duo T7200 at 2.0 GHz with 1 GB 667 MHz PC2-5300 DDR2 SDRAM memory) running Windows XP Version 5.1 Service Pack 3. We explicitly did not install any antivirus software on the machine since we knew that some of them automatically nest themselves into several of the browsers we are testing to prevent phishing attacks. These antivirus browser plug-ins might affect a browser performance in the time it needs to render a webpage hence we opted not to install any antivirus software.

The MacBook we use only has 13.3" LCD Widescreen with a native resolution of 1280 by 800 pixels which isn't sufficient for our experiments since we plan to use 1280 by 720 pixels photos in our experiments. We therefore have attached a 24 inch Dell 2407 WFPb running at a resolution of 1920 x 1200 @ 60Hz which has been set up as our primary monitor. The choice for this particular Apple Mac was made since we believe it comes very close to the average PC a normal consumer is using nowadays to browse the World Wide Web.

Our backend server powering the web application was placed within the LIACS Media Lab at Leiden University and our Apple is connecting to it via gigabit network to a fibre line capable of delivering a 100Mbits. The quality and speed of the connection was of such a high level that in a preparation test we discovered that the load times of the same image were virtually constant.

### 4.1.2 Web browsers

Browser vendors are releasing new versions of their product almost every month and each iteration usually incorporates an improvement of overall performance and better support of web standards. We would like to caution our readers that this frequent cycle of software updates causes our experimental results to be only valid for a couple of months.

All browser tests in this thesis were preformed with the latest version that was available on the vendor website and which was labeled stable and released with the intention to be used by the general public in August 2009. Furthermore no additional plug-ins or extensions were installed besides the ones that were included by default within the original download.

If we discuss a particular browser within this thesis the following browser version should be filled in:

- Apple Safari 4.0.2.530.19.1
- Opera 9.64.10487
- Microsoft Internet Explorer 8.0.6.001.18702
- Mozilla Firefox 3.5.2 (Gecko 200090729)

During the preparation of the experiments we noticed that most browsers have poor memory management. The so-called garbage collector in every browser should free up memory if it detects that previous allocated memory is not in use anymore after which it is released back to the operating system.

We however found out that especially Mozilla Firefox does not promptly return the memory it needs to display a webpage after closing it and navigating to another one. Instead it looks like Firefox keeps on consuming memory until a certain predetermined point and then starts unloading the memory it does not need anymore. Since we have planned an experiment in which we measure memory utilization behavior we decided to do all our tests with a 'cold boot'. Cold booting means we start every test by ensuring that there are no running instances of a particular browser.

In a second step we use a little piece of additional software called CCleaner from Piriform Ltd. to permanently remove any files on the hard drive that contain data of previous browser sessions. By deleting the browser history, cookies and cache we make sure that every test has the same starting conditions which thus include a minimal usage of memory.

## 4.2 Measuring application times

The first experiment targets the usability of applying web graphics technology to perform browser-based image manipulation. Simply put we want to know how long it takes between a user giving the order to manipulate an image and that the result is shown back to him. For the experiment we have written a special set of HTML files that directly call our JavaScript web graphics rendering engine, the rest of our application has been left out for simplicity and performance reasons.

Our aim of this experiment is to see whether application of an image manipulation operation can be done within a reasonable time. A quick study of related literature [43] shows there conflict in evidence from which we concluded that the response time should preferably be less than 2 seconds (Scheiderman 1984) or at max 4 seconds (Galletta, Henry, McCoy and Polak 2004) Any time longer over 4 seconds means we would have to incorporate a 'please wait' notification message into the application to inform the user that we have acknowledged his order to alter the image. Application times of greater than 30 seconds or an unresponsive browser than it might be better to stick with the traditional server-based image manipulation.

Besides determining the usability we are also interested in the relation of the size of the image and the application time. Therefore we have selected two 1920 x 1200 images; one (image B) with a relative big 1.17 MB file size contains a high level of colors and details. The second photo (image F) on the other hand is only 305 KB and has hardly any colors or details. We then resized and cropped both images to the following formats: 480 x 360, 576 x 384, 900 x 600 and 1280 x 720 pixels.

This a odd set of formats is created by design; the 576 x 384 and 900 x 600 resolutions are the result of taking the classic standard 3:2 photo format of 6 by 4 inches and multiplying it with the dots-per-inch setting of Windows XP and the lowest acceptable quality of our printer which is respectively 96 and 150 DPI. The 480 x 360 and 1280 x 720 resolutions are popular 16:9 formats for displaying video on the web and were hence chosen.

Our test HTML webpage will show one of the two images on load of the page and four buttons allowing to rotate 70°, grayscale, invert the colors or to crop the photo to a 300 x 300 pixels starting at 50,50 in the original image. On press of a button with the mouse we use JavaScript's *dateObject.getTime* method to get the current time in milliseconds since January 1$^{st}$ 1970 and save it to a variable. On completion of the application of one of the operations we again fetch the current time after which we calculate elapsed time in milliseconds and using a JavaScript's *Alert* method we display it onto the screen. We will perform and measure every operation twice and then calculate the average application time.

### 4.3   Level of interactivity

Being able to manipulate an image within the browser is great but we are set on the idea to make working with photos online more interactive. We have implemented drag & drop, rotation and scaling using a mouse but we need to make sure that this will work smoothly for all our potential end-users. We therefore have devised a test to measure how well photos rendered with Canvas, SVG and VML can keep up with the user's mouse actions. Again we have created a custom HTML file that only included and uses the rendering library and loads a single photo. For our convenience we will use the same set of images we have created for the previous test.

Using a small piece of software we have recorded a 2681 milliseconds 503 pixels downwards mouse movement. We then apply this movement using the same software to the rendered photo of our HTML test page causing the photo to move down the page. By repeating the mouse move over and over for every image size we will be able to compare the different web graphics technologies by time it takes to render the movement.
In order to measure the time it takes we do not rely on a stopwatch or use JavaScript's *dateObject.getTime* method but we thought it was a better idea to use a camera to record every move in all the different browsers and then use frame-by-frame playback software to determine the exact times, CPU utilization and memory usage.

The use of a 25 frames per second video camera will give use more accurate results than the typical human reaction time of approximately 150 to 300 milliseconds. However 25 frames per second comes down to one frame every 40 milliseconds which means that in determining the exact start & end frame we will have an margin of 80 milliseconds resulting in a $80/2681 * 100 \approx 3\ \%$ error window. This quite a significant error but our goal is not to determine accurate times down to the millisecond but just to find out at which image size and browser a technology will start to be so out of sync with the mouse movement that it will be noticeable by the user.

Another item that we will be measuring is the CPU utilization and memory usage of the browser since these numbers will give us an insight to the amount of resources that are needed to run our web application. After the experiment we should than be able to extrapolate the results so we are able to determine the minimal system requirements. A secondary use of the numbers will be to spot possible bottlenecks in our application e.g. huge CPU spicks which need to be resolved in future releases in order to make the application work better on low end PCs.

Canvas fundamentally works different in how it renders graphics in comparison to SVG and VML. Obviously rendering an entire canvas multiple times per second with JavaScript is more performance heavy than moving a DOM element using browser functions. Applying an effect to a photo should even increases the CPU usage and since we allow the user to grayscale, change brightness, hue, saturation or invert the colors of the photo we should include a movement of manipulated photos in this experiment. Due to limited time and resources we have chosen not to test all implemented effects but to restrict ourselves to testing with photos where the colors were inverted using Canvas, SVG or VML.

## 4.4 Simulating real-life usage

All previous experiments are all only using one single photo but at this point we would like to know how well our web application performs when we start loading it up with multiple photos. In this last test we are going to simulate actual usage by an end user by continuously loading up multiple photos and then perform and measure a rotation operation every 5 photos. The aim of this experiment is to find out the maximum amount of photos a browser using a particular rendering technique can handle in real-life.

We have created a set of 100 576 x 384 and 100 900 x 600 JPG photos with a file size that varies between 9 and 376 KB and an average file size of 109 KB. In order to simulate real-life usage we have written a script that depending on how many images have already been loaded performs a manipulation operation onto the photo.

The pseudo code below shows how we simulated.

```
if (not divisible by 5) {
   Set position of photo o random location within browser window
} else {
   Position of the photo is in the center of browser window
}

if (divisible by 2) {
   Choose 576 by 384 pixels photo source image
} else {
   Take a 900 by 600 pixels photo as source image
}

if (divisible by 3 and not divisible by 5) {
   if (divisible by 2) {
      Rotate the image negative 50 degrees
   } else {
    Perform a 10 degree rotation
   }
} else if (divisible by 5) {
   Scale the photo to 80 percent of its original dimensions
}

if (divisible by 7) {
   Crop to image to a 300 by 300 pixels photo starting
   at point 50,50 in the source photo
}
```

Finally we wrote up a set of stop rules that will be used to determine the usable upper limit of the number of photos a particular technology with a specific browser can handle.

- A new photo must appear within 15 seconds after clicking the button. This is quite reasonable seeing the average file size of 109 KB and the capacity or our internet connection.
- An warning message of unresponsive or slow script appears.
- The browser is unresponsive for more than 30 seconds or crashes.
- 100 photos have been loaded onto the webpage.

The results of this test will show the robustness of Canvas, SVG & VML and will help us set a save upper limit of the number of photos that can be displayed on the same page within our web application.

# 5. Discussion

## 5.1 Application times

### 5.1.1 The crop & rotate operation

Let us start off with tables 2 and 3 which show the average application times we recorded when performing a crop starting at point 50, 50 in the original photo with a width of 300 pixels and a height of 300 pixels and a rotate operation. Everyone can see that the Canvas times are significantly higher than the SVG and VML. This was to be expected since Canvas needs to re-render the entire drawing canvas which is a costly operation.

| Images | VML IE 8.0 | Canvas FF 3.5.2 | SVG FF 3.5.2 | Canvas Safari 4.0.2 | SVG Safari 4.0.2 | Canvas Opera 9.64 | SVG Opera 9.64 |
|---|---|---|---|---|---|---|---|
| b480x360 | 0 | 246.5 | 1.5 | 456.5 | 4.5 | 2124.5 | 0 |
| b576x384 | 0 | 309 | 1.5 | 574.5 | 4 | 2742.5 | 0 |
| b900x600 | 0 | 793 | 1 | 1444.5 | 4 | 7102 | 0 |
| b1280x720 | 0 | 1263.5 | 1 | 2555 | 4.5 | 12398 | 0 |
| f480x360 | 0 | 243.5 | 1 | 447 | 4 | 2156.5 | 8 |
| f576x384 | 0 | 306.5 | 1.5 | 576.5 | 4 | 2774 | 0 |
| f900x600 | 0 | 747 | 1.5 | 1463.5 | 4.5 | 7016 | 0 |
| f1280x720 | 0 | 1293 | 1.5 | 2557 | 4 | 12242.5 | 0 |

**Table 2.** Average application times in milliseconds of a crop operation (50, 50, 300, 300)

The second trend that stands out is string of zero values in the VML Internet Explorer and the SVG Opera columns. At first we thought we had made an error in our code thinking that the *dateObject.getTime* method was not 100% cross browser properly supported.

We started to read the browser's documentations and search online for the answer which revealed to use something unexpected. We learned that Internet Explorer, Opera and Safari browsers are only updating their internal *dateObject.getTime* representations on average every 15 milliseconds. Any test that takes less than 15 milliseconds will therefore always be rounded down to 0 milliseconds in these browsers making it impossible to get accurate times. Luckily for us we only were measuring application times in order to make sure that they were less than 4000 milliseconds at max. If we look at tables 2 & 3 only Canvas in combination with the Opera browser is surpassing that threshold.

| Images | VML IE 8.0 | Canvas FF 3.5.2 | SVG FF 3.5.2 | Canvas Safari 4.0.2 | SVG Safari 4.0.2 | Canvas Opera 9.64 | SVG Opera 9.64 |
|---|---|---|---|---|---|---|---|
| b480x360 | 0 | 465.5 | 1.5 | 896.5 | 2 | 4125.5 | 0 |
| b576x384 | 0 | 602 | 1 | 1143.5 | 3 | 5430 | 0 |
| b900x600 | 0 | 1400 | 1 | 2640.5 | 3 | 12758 | 0 |
| b1280x720 | 0 | 1816.5 | 1.5 | 3374.5 | 3 | 16023.5 | 0 |
| f480x360 | 0 | 479.5 | 2 | 890 | 3 | 4156 | 0 |
| f576x384 | 0 | 622.5 | 1.5 | 1148.5 | 3 | 5390.5 | 0 |
| f900x600 | 0 | 1415 | 1 | 2646 | 2.5 | 12600.5 | 0 |
| f1280x720 | 0 | 1781.5 | 1.5 | 3439.5 | 3 | 16094 | 0 |

**Table 3.** Average application times in milliseconds of applying a 70° rotation to a photo.

### 5.1.2 Grayscaling and inverting the colours

Moving onto the grayscale and inverted colour operation we only have to look at the numbers in tables 4 & 5 that are over 15 milliseconds. Again we see that Canvas is well over the 4000 milliseconds benchmark which means that when Canvas is chosen as the rendering technology we really need to insert a notification message when a user applies an effect. The b1280x720 contains three 'not applicable' values due to the fact that during the experiment the browsers became unresponsive for longer than 2 minutes and hence no time was recorded.

| Images | VML IE 8.0 | Canvas FF 3.5.2 | SVG FF 3.5.2 | Canvas Safari 4.0.2 | Canvas Opera 9.64 | SVG Opera 9.64 |
|---|---|---|---|---|---|---|
| b480x360 | 0 | 415.5 | 2.5 | 778 | 3164 | 0 |
| b576x384 | 8 | 539.5 | 2.5 | 967.5 | 4101 | 0 |
| b900x600 | 0 | 1380.5 | 2.5 | 2356.5 | 10023.5 | 0 |
| b1280x720 | 0 | n/a | 2.5 | n/a | n/a | 0 |
| f480x360 | 0 | 420 | 3 | 772 | 3179.5 | 0 |
| f576x384 | 0 | 544 | 2 | 972 | 4117 | 0 |
| f900x600 | 0 | 1297.5 | 2.5 | 2357 | 9937 | 0 |
| f1280x720 | 0 | 2219.5 | 2 | 4058.5 | 17132.5 | 0 |

**Table 4**. Average application times of applying a grayscale effect to a photo.

| Images | VML IE 8.0 | Canvas FF 3.5.2 | SVG FF 3.5.2 | Canvas Safari 4.0.2 | Canvas Opera 9.64 | SVG Opera 9.64 |
|---|---|---|---|---|---|---|
| b480x360 | 0 | 508.5 | 2.5 | 978.5 | 4.5 | 0 |
| b576x384 | 8 | 651 | 3 | 1200.5 | 4 | 0 |
| b900x600 | 8 | 1580.5 | 3 | 2985.5 | 4 | 0 |
| b1280x720 | 0 | 2673 | 2.5 | 4980.5 | 4.5 | 0 |
| f480x360 | 0 | 524 | 3 | 959 | 4 | 0 |
| f576x384 | 0 | 652 | 2 | 1200.5 | 4 | 0 |
| f900x600 | 0 | 1564 | 3 | 2921 | 4.5 | 0 |
| f1280x720 | 0 | 2640.5 | 2 | 5072 | 4 | 0 |

**Table 5.** Average application times of inverting the colours within a photo.

## 5.2 Interactivity results

### 5.2.1 Results of measuring mouse movement times

In figure 7 have plotted the measured delta times to the 2681 milliseconds scripted mouse move against the different image size where every line stands for a particular browser. The graph clearly shows SVG and VML are faster in handling the mouse interaction than Canvas.

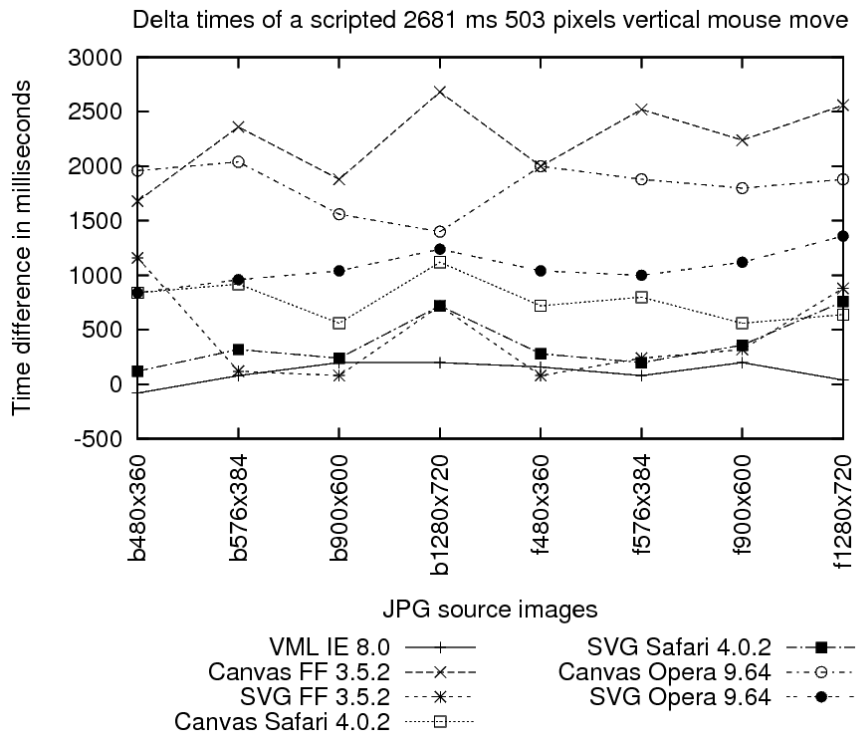

**Figure 7.** Measured delta movement times against a scripted mouse move of a photo

This isn't that surprising since for every update of the mouse position Canvas redraws the entire scene. Still when we played back the videos we only could detect a slightly noticeable stuttering in the two upper lines e.g. Canvas on Firefox and Opera who have a delay of almost 2 seconds against the original mouse move.

The relation between the size of the image and its move time is showing in the steadily increasing line whereas the level of detail e.g. whether photo B of F was used does not seems to have any effect. There are here and there quite some anomalies in the graph see for instance the dip at b1280x720 in the Canvas Opera line which we cannot logically explain and should therefore be written down to experimental errors.

### 5.2.2 Processor utilization numbers

Figure 8 shows the average CPU utilization which we recorded during the experiment in order to determine the minimal system requirements of our web application. The lowest overall CPU usage is on the name of Microsoft's Internet Explorer although one should point out that this browser is heavily integrated with the underlying operating system which is of the same vendor. Therefore it is no more than likely that parts of the workload of the rendering the web page was shipped off to other operating system processes which would explain the low CPU numbers since we only measured the browser process(es).

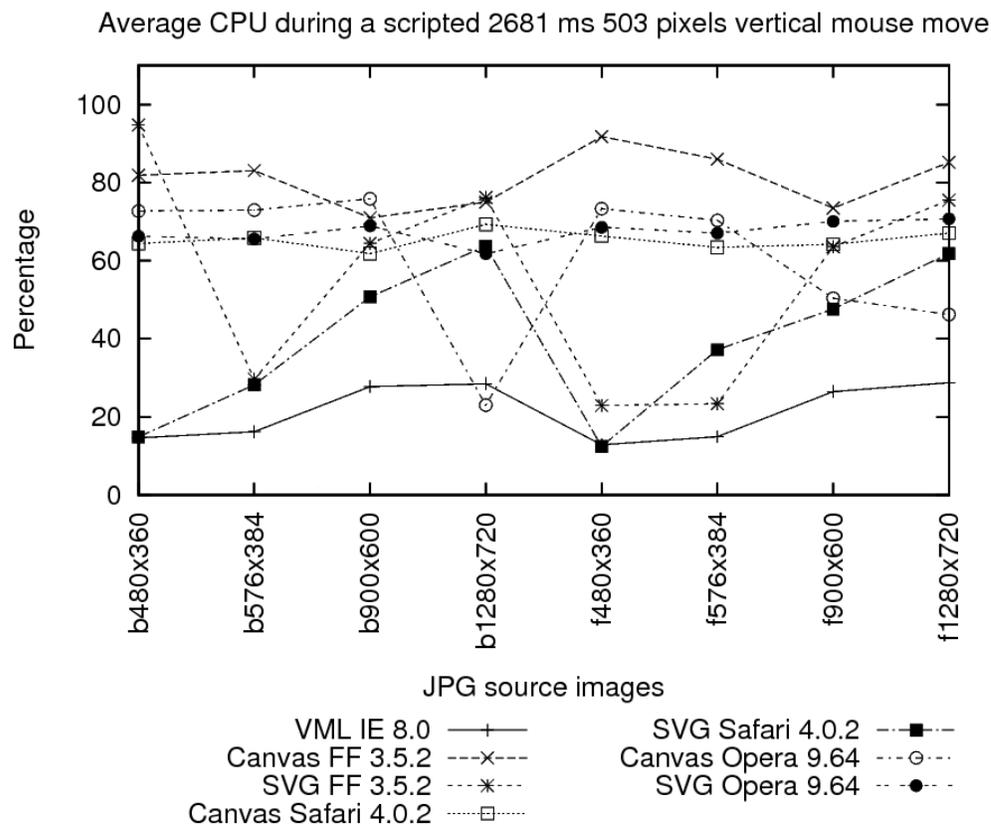

**Figure 8.** Average CPU utilization during a scripted mouse move of a photo.

The figure on the previous page clearly shows that the majority of the Canvas and SVG lines are in the 60-80% average CPU utilization meaning that our application is really on the heavy side for a single web application. Our test machine is from the year 2006 so we expect that an end-user will need to have a PC not older than 4 year in order to run our prototype web application without any hiccups using Canvas or SVG.

The second range of tests which we preformed in the same experiment was to look at the impact of applying the colour inverted effect on the mouse movement times and resource usage.

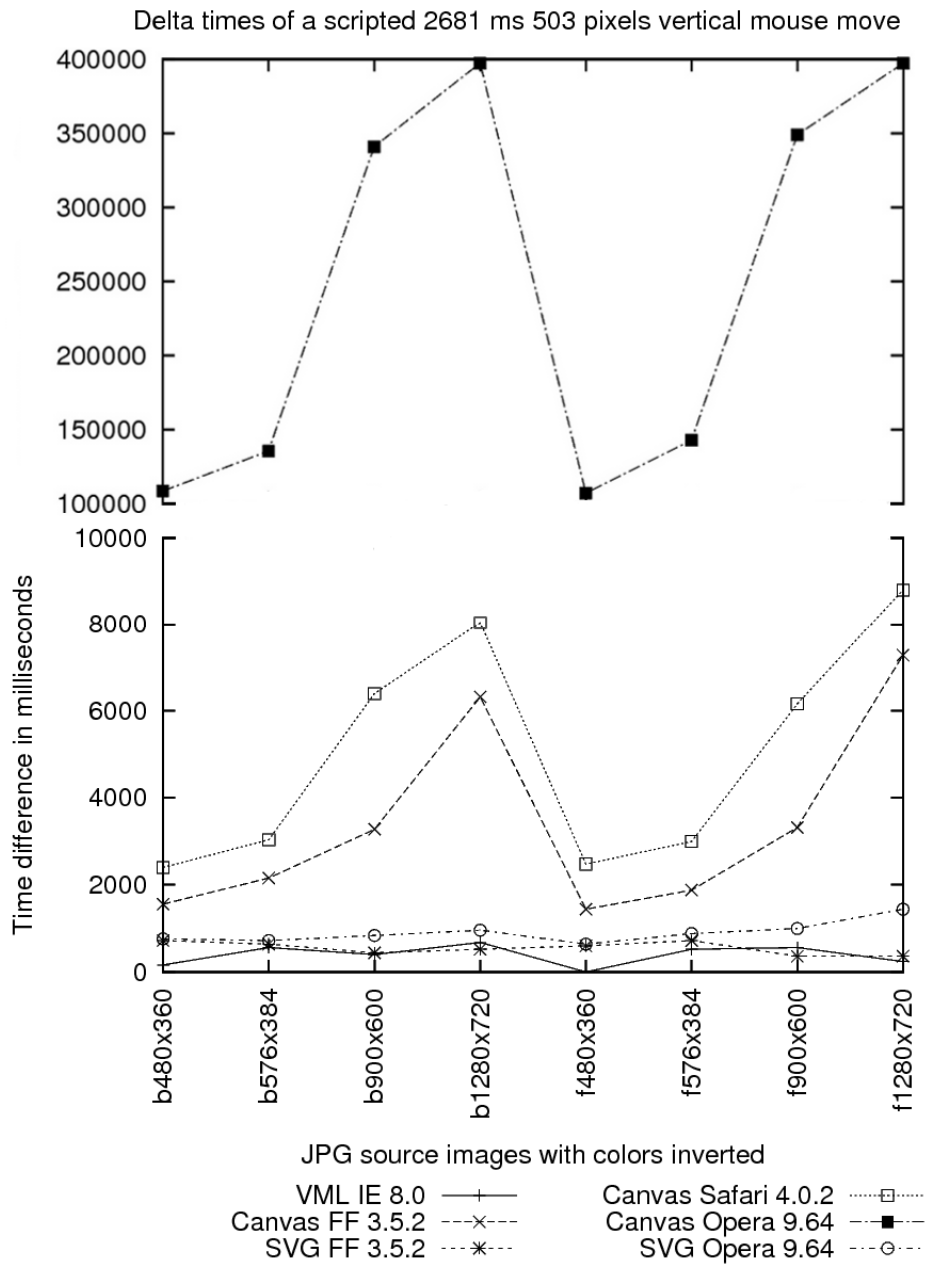

**Figure 9.** Delta movement times of a scripted photo mouse move with the colours inverted.

Figure 9 displays that Canvas is current not the technology to use when designing highly interactive web graphics. During the experiment rendering the movement of any images using Canvas was something that was really pushing against the edges of the capabilities of the browser. In Apple's Safari the original photo would be shown and then at the last moment it jumped to its final destination and then the inverted effect would be applied. Firefox and Opera showed a slowly heavily stuttering photo where only parts of the photo had been inverted if Canvas was used. SVG and VML on the contrary had no problem keeping the photo in sync with the mouse and showed no noticeable delays.

We have plotted the CPU utilization belong to the movement of a photo with inverted effect in figure 10. It was expected that the average CPU number for Canvas would be higher and they are except for Opera which is showing almost the same numbers as in figure 8. SVG and VML numbers are about 20 % higher than when rendering the original images but a second look at figure 9 shows that the majority of the SVG and VML times are all in the same range and are near similar to moving a non manipulated photo (figure 7). This means that application of an effect on interactive SVG or VML DOM elements will spike CPU usage considerably but does not seem to affect overall usability.

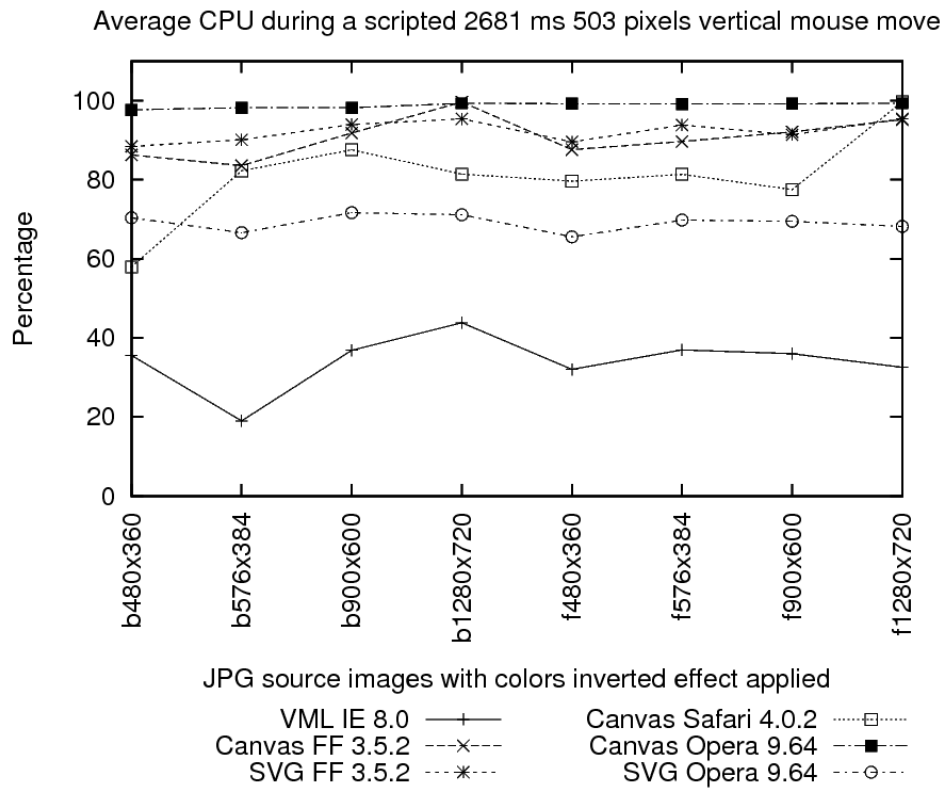

**Figure 10.** Average CPU utilization during a mouse moveof a photo with the colours inverted.

### 5.2.3 Memory usage during the experiment

A third factor that we measured during this experiment was memory usage and we have plotted the changes in memory usages in figures 11 & 12.

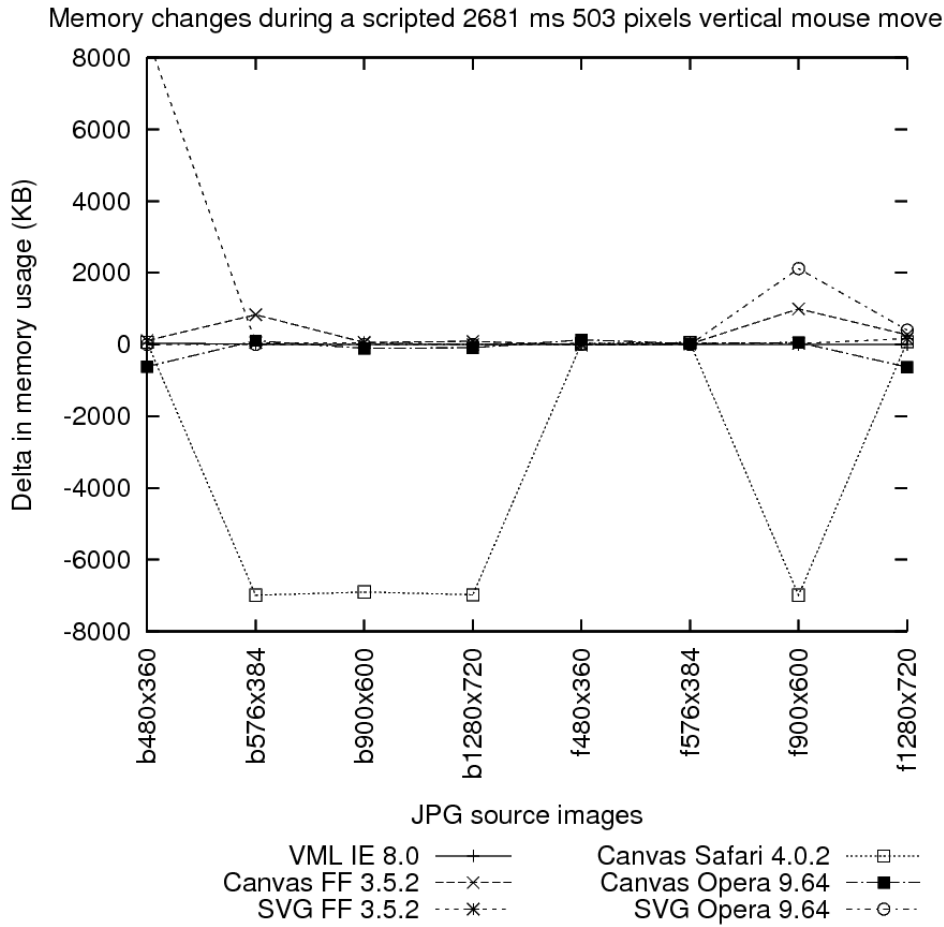

**Figure 11.** Fluctuations in memory usage during the move experiment

Overall memory consumption during the experiment was relatively constant for the majority of the browsers but two trends did pop out. First of all it looks like Apple has built in a memory optimization technique in Safari when Canvas has to quickly render the same scene multiple times. Both graphs show a drop in the amount of memory used after a movement of the photo.

The second visible trend in both figures is completely the opposite; Firefox and Opera have a clear tendency to consume vast amount of memory when they have trouble rendering a <canvas> element. Explaining both trends is not that difficult; Apple originally designed Canvas and has therefore a head start on the Firefox and Opera development teams.

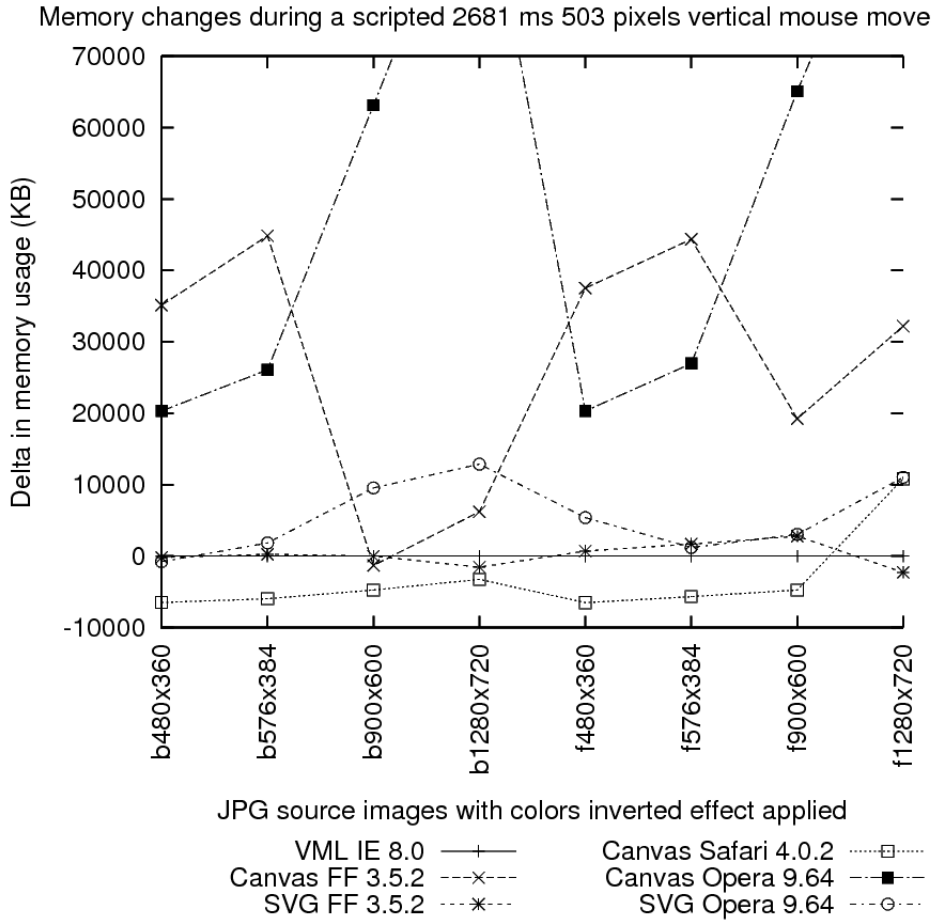

**Figure 12.** Memory fluctuations during mouse move with colours inverted effect applied.

### 5.3  Web graphics in real-life

In our last experiment we were trying to simulate real-life usage of our web application in order to find out the maximum number of photos the average web browser can handle using the various web graphics rendering technologies.

Figure 13 shows the times it took rotating a photo negative 111.8 degrees with an ever increasing number of photos already loaded. The times are significantly increasing around the 35 – 40 loaded photos this is exactly the same range in which we started to notice during the experiment that photos were not moving in sync with the mouse anymore.

The save upper limit should therefore be set at 30 mark to keep the users experience smoothly. This number however only applies to SVG and VML since although hardly visible in figure 13 all Canvas runs failed between the 7 and 17 loaded photos. The Opera browser was not able to load photo number 7 in under 30 seconds; Safari gave a slow script warning at 8 loaded photos. Firefox gave an unresponsive script warning message at 17 which again confirms our earlier conclusion that Canvas is currently not suited for rendering highly interactive web pages.

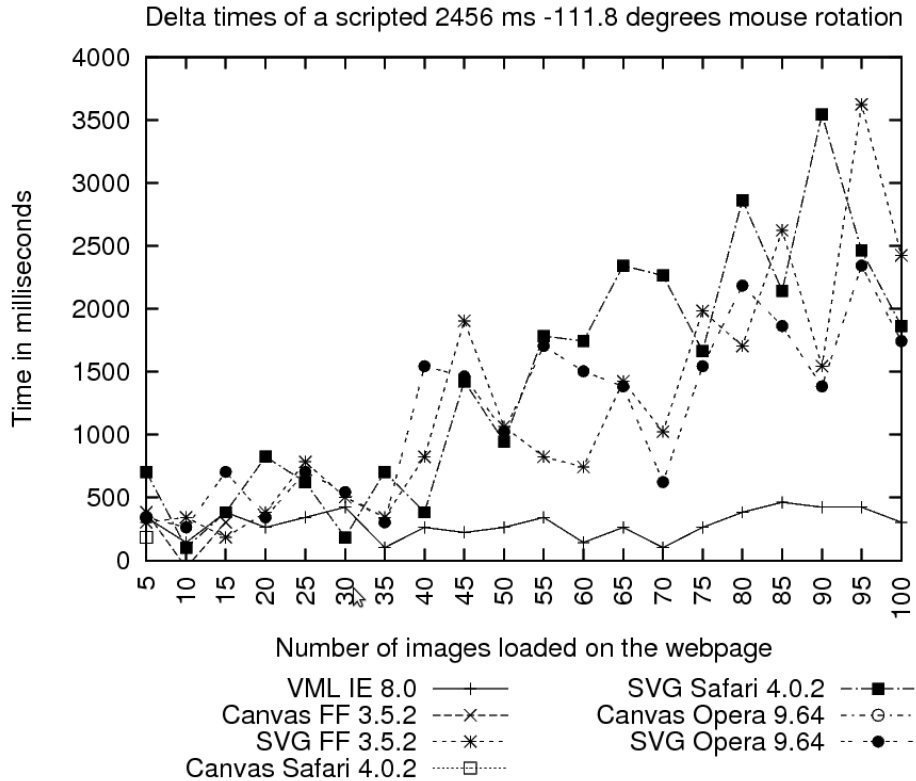

**Figure 13.** Rotation times of a single photo when multiple photos on the page.

## 5.4 Looking at the numbers

There are a couple of additional remarks that should be placed on the results of our second experiment in chapter 5.2. The more careful reader probably already had noticed that the numbers for greyscale and colours inverted for SVG on Safari are missing. The reason behind this is very practical due to the fact that Apple did not in corporate SVG effects in their browser. Secondly in figure 9 the numbers of the Canvas Opera for b1280x720 and f1280x720 have been fixed at 400000 since at that point in time we stopped the measurement. The version of Opera that we tested was just not capable of performing the desired movement operation with a large image with the colours inverted effect applied within a reasonable timeframe. Thirdly all of the average Canvas Opera CPU numbers in figure 10 were not calculated as the rest of the numbers but were guessed based on playback of the video. We had to resort to this solution since calculation of the average CPU by watching frame-by-frame and then calculating the average CPU was an immense and impossible task giving that times were ranging between 100000 and 400000 milliseconds

Overall the data we retrieved from our experiments shows that SVG and VML are capable to perform the desired photo manipulations in sync with the user actions up to a limit of around 35 photos. All of our experiments demonstrated that Canvas, although very capable in theory, has yet to be optimized in all of the tested browsers to unleash its manipulation capabilities in a useable manner to an end user. We would therefore recommend the use of SVG in combination with VML if one is developing a web application for the general public with a release date of less than 6 months.

# 6. Conclusions and future works

## 6.1 The viability of browser-native image manipulation

In this thesis we presented a prototype web application that makes it possible to perform image manipulation operations using browser-native technology. By only getting a reasonable numbers of image manipulation operations to work we already partly answered our research question whether it was technically possible. Our experiments have shown that there are some restrictions on the size and number of images that can be loaded on a single web page using web graphics. Overall web graphics technology at this moment in time is capable to crop, move, rotate and scale multiple photos albeit that developing using JavaScript the software is not a walk in the park.

Actual deployment of the technology for use by a broad public is nonetheless not possible just yet. The main roadblock for widespread adaption is in our eyes still Microsoft's Internet Explorer with a market share of around 60 percent which does not support Canvas or SVG. Internet Explorer's VML is an old, antiquated and almost dead proprietary technology which is not capable of doing the more advanced effects such as changing hue and saturation or even a simple sepia effect. Hopefully the next incarnation of Internet Explorer will be fully HTML 5 and thereby supporting the <canvas> element. Canvas on its own is currently only sufficient for displaying relative small and static objects. If designing a more interactive user interface we recommend SVG should be used for handling dragging and rotation of objects and Canvas for changing little details and effects. We however expect that Canvas will be more powerful and faster than SVG in the near future since several teams are currently working on OpenGL powered rendering of Canvas. The Canvas 3D JavaScript Library [43] is using the power of the graphics card to render Canvas content and although in its infancy their early performance numbers are quite astonishing.

At this point in time we are able to release a web application to the world that changes the way people are allowed to interact with photos online. We have shown that it is possible to get rid of the grid lay-out and move on to an interface similar to a traditional scrapbook.

## 6.2 Future works

Since browser development happens at a rapid pace it would be nice to perform the same set of experiments in about a year. This would allow us to independently verify vendor claims of how well the performance of their browser has been increased since the previous version. Our experiments revealed that measuring browser performance using JavaScript's *getTime* functions is inaccurate and we found a solution to this problem by using a video camera. If we did the experiments again we would have make some changes to our current setup foremost we would use a 100 or 120 Hz screen and a high-speed video camera in order to reduce our error window to less than 10 milliseconds.

The interface for interaction with the photos is already almost perfect for the use with touch technology. In order to make our web application to work on for instance a multi-touch phone like the Apple's IPhone all that needs to be done is write a small JavaScript layer that hooks up touch commands to our rendering engine. We believe that touch technology in combination with web applications will quickly became common place since Apple, Microsoft and Mozilla have all demonstrated touch version of their respective browsers. This shift in interface will open up a whole new way people of how are interacting with their photos on the World Wide Web.